\documentclass{elsarticle}
\usepackage{lineno,hyperref}
\usepackage{gensymb}
\usepackage{amsmath,amssymb,todonotes}
\modulolinenumbers[5]
\DeclareMathOperator{\trace}{tr}
\DeclareMathOperator{\deviatoric}{dev}
\journal{Computer Methods in Applied Mechanics and Engineering}
\usepackage{multicol} 
\usepackage{changes}
\usepackage{todonotes}
\usepackage[ruled]{algorithm2e}
\usepackage{graphicx}
\usepackage{subfig}
\usepackage{graphicx}
\usepackage{numcompress}
\usepackage{booktabs}
\usepackage{algorithmic}
\usepackage{appendix}
\usepackage{upgreek}
\begin{document}

\begin{frontmatter}

\title{A ``parallel universe'' scheme for crack nucleation in the phase field approach to fracture}

\author[mymainaddress]{Yihao Chen}

\author[mymainaddress,myotheraddress]{Yongxing Shen\corref{mycorrespondingauthor}}
\cortext[mycorrespondingauthor]{Corresponding author}

\address[mymainaddress]{University of Michigan -- Shanghai Jiao Tong University Joint Institute, Shanghai Jiao Tong University, Shanghai, 200240, China}
\address[myotheraddress]{Shanghai Key Laboratory for Digital Maintenance of Buildings and Infrastructure, Shanghai, China, 200240}

\begin{abstract}
     Crack nucleation is crucial in many industrial applications.
    The phase field method for fracture transforms the crack nucleation problem into a minimization problem of the sum of the elastic potential energy and the crack surface energy. Due to the polyconvexity of the formulation, starting from a crackless solid, a standard Newton iteration may lead to a solution with no crack, even though a cracked solution has a lower total energy. As such, the critical load for cracking is highly overestimated. Here, we propose an algorithm termed ``parallel universe'' algorithm to capture the global minimum. This algorithm has two key ingredients: (a) a necessary condition for cracking solely based on the current crackless solution, and (b) beginning from when this condition is met, Newton iteration with two initial guesses, a crackles one and a cracked one, will both be performed and the converged candidate solution with lower energy is accepted as the solution at that load step. Once the cracked candidate solution is accepted, the crackless one is discarded, i.e., only one universe is retained. This cracked initial guess is obtained only once for all load steps by solving a series of similar minimization problems with a progressively reduced critical crack energy release rate. Numerical examples with isotropic and anisotropic critical crack energy release rates indicate that the proposed algorithm is more reliable (as there is no need to retrace) and more efficient than the standard Newton iteration and a well-known backtracking algorithm.
\end{abstract}


\end{frontmatter}


\section{Introduction}
Crack nucleation is crucial in the modeling of many processes with technological significance. 
One theory that addresses the crack nucleation problem of brittle materials is the variational thoery of fracture put forth by Francfort and Marigo \cite{FRANCFORT19981319}. 
A regularization of this theory with a length scale parameter $\ell$ was proposed by 
 Bourdin et al.~\cite{BOURDIN2000797}, which permits efficient implementation and which later adopts the name the phase field approach to fracture.
The phase field method has become one of the mainstream methods for fracture simulation. 

While this method yields satisfactory results for problems with pre-existing cracks, how to predict crack nucleation in general and in the case of fracture phase field remains a challenge especially when the domain, the load, and the material are all homogeneous.

As is well known, when there exist multiple local minimizers for the energy functional, the solution given by the Newton method is not always the global minimizer, but often a local minimizer close to the initial guess. Therefore, starting from a crackless solid, even though the applied load reaches a certain level such that a cracked solution gives a lower total energy, a standard Newton iteration may still lead to a solution with no crack, until when the applied load is excessively large such that the strain energy dominates and any crackless initial guess leads to a cracked solution. As a result, the critical load for cracking is highly overestimated. This has a profound consequence, as it leads to the need for a high safety factor for design purposes. 

A couple of approaches have been proposed to tackle this problem. A notable example with the potential of converging to the global minimizer is the backtracking algorithm proposed by Bourdin \cite{BOURDIN2007411}.
 In problems with proportional displacement loading, this algorithm samples more candidates in addition to those obtained from the standard Newton iteration by scaling newly obtained solutions to the loads of previous steps, increasing the chance of finding the global minimizer. Consequently, it is very likely that the provisional result in each load step is modified by later steps. Therefore, much more computation after the desired final load is needed in order not to miss better candidates.
 
As another example, Kopaničáková and Krause \cite{KOPANICAKOVA2020112720} developed a recursive multilevel trust region method (RMTR) to address this minimization problem. The authors combine the trust region method with the multilevel method to accelerate the monolithic solution process. In particular, they employ level-dependent objective functions for minimization. This RMTR method is shown to be much faster than the normal trust region method and the staggered solution scheme. Nevertheless, if the distance between the global minimizer and the current guess is larger than the trust region step size, a similar situation as the standard Newton iteration may occur, i.e., the algorithm may still converge to a local minimizer.


In principle, the crack nucleation problem at hand can be solved with methods designed for general global minimization problems, such as simulated annealing \cite{Kirkpatrick1983} and the genetic algorithm \cite{fraser1957simulation}. While these are standardized procedures, it remains a challenge how to incorporate the physics into the problem so that a big sampling space (say, all phase field degrees of freedom) can be avoided. 

In this work, we propose a ``parallel universe" scheme, aiming to address the crack nucleation problem. The idea is to first find a cracked initial guess when a certain criterion is met. Then we re-solve the problem with this cracked initial guess. Then normally we have both a crackless candidate solution and a cracked one. We then label the candidate solution with a lower total energy as \emph{the} solution for the current load, but both candidates will be retained for subsequent calculations as initial guesses, and hence the name ``parallel universe.'' In most cases the crackless candidate yields a lower total energy. Once this relation is reversed, the crackless candidate will be discarded, due to irreversibility.

The proposed scheme is efficient in two aspects. First, when the said criterion is not met, only the crackless candidate needs to be tracked. Second, although the process to find the cracked initial guess is relatively expensive, once that is available, the critical load for cracking (that is, the critical load such that the cracked candidate has a lower total energy) will not be missed. Moreover, there is no need to compute for a load level higher than the desired one, as opposed to the case of the backtracking algorithm.

The structure of this work is as follows. Section \ref{sec:ProbState} states the problem in more details and briefly describes the proposed scheme. On this basis, Section \ref{sec:Method} introduces the basic phase field formulations and details necessary for the proposed scheme, and finally the entire scheme. The proposed scheme is verified with numerical examples in Section \ref{sec:Numerical}. According to the results in Section \ref{sec:Numerical}, the proposed scheme is compared with the standard Newton method and the backtracking method in the aspect of the computational time and accuracy.


\section{The scheme at a glance}\label{sec:ProbState}
\subsection{Problem statement}
To concentrate on the main idea, consider a solid which may undergo brittle fracture. For simplicity, the solid is assumed to be under only displacement loading but no traction or body force is present. Let $\Omega\subset \mathbb{R}^{n}$, $n=1,2,3$, be a regular-shaped bounded domain occupied by the solid in its undeformed state. This energy functional of $\Pi$ is given by: 
\begin{equation}
\Pi [\boldsymbol{u},d]=\int_{\Omega}\psi(\nabla\boldsymbol{u},d)\,\mathrm{d}\Omega +\int_{\Omega} G_c\left(\frac{d^{2}+\ell^{2}\left| \nabla d\right|^{2} }{\mathrm{2} \ell}\right)\mathrm{d}\Omega,
\label{Energy equation}
\end{equation}
where $\psi$ is the strain energy density, $\boldsymbol{u}:\Omega\rightarrow\mathbb{R}^n$ is the displacement field and $d:\Omega\rightarrow[0,1]$ is the phase field. Parameters $G_c>0$ and $\ell>0$ are the critical crack energy release rate and the phase field length scale parameter, respectively. The phase field approach is formulated as to find a global minimizer of $\Pi$ subject to the constraints $0\le d\le 1$ almost everywhere in $\Omega$ and the displacement boundary condition (load) $\boldsymbol{u}=\boldsymbol{u}_b$ on $\Gamma_D\subseteq\partial\Omega$.

The fields $\boldsymbol{u}$ and $d$ can be solved either in a monolithic way or by alternating minimization. In both cases, the solution process is usually based on the Newton method. More specifically, the initial guess for the Newton iteration in each step is the solution of the previous step, except for nodes with prescribed displacements. 

At this point, it will be useful to restrict ourselves to a widely adopted form for $\psi$, i.e., $\psi=(1-d)^2\psi_+(\nabla\boldsymbol{u})+\psi_-(\nabla\boldsymbol{u})$. For now, it is sufficient to know $\psi_\pm$ are quadratic functions of $\nabla\boldsymbol{u}$ and $\psi_\pm\ge0$.

\begin{figure}[htbp]
   \centering
   \includegraphics[width=12cm,keepaspectratio]{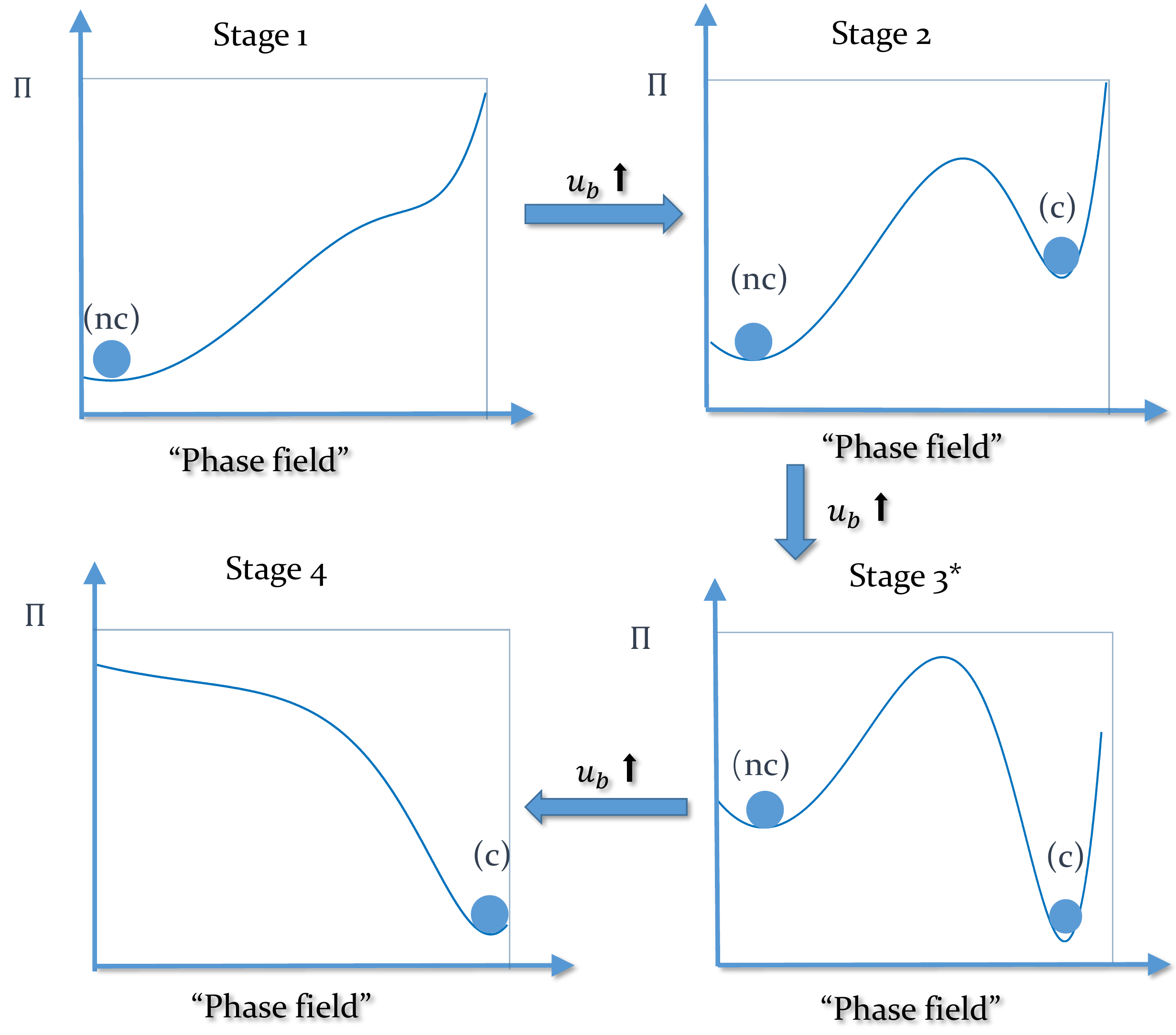}
    \caption{Illustration of the difficulty of the standard Newton iteration for the crack nucleation problem in the phase field approach for crack nucleation. The solution process from the beginning of loading to ultimate fracture is represented in a few stages. In each subfigure, the horizontal axis represents a certain macro-coordinate for the phase field $d$ for illustration purposes. Symbols (c) and (nc) represent local minimizers with and without a crack, respectively.}
    \label{Diagram}
\end{figure}

Figure \ref{Diagram} illustrates the solution process of a crack nucleation problem from the beginning. In Stage 1, the load $\boldsymbol{u}_b$ is small enough so a crackeless result (nc) is the only local (and global) minimizer. In Stage 2, there exist two local minimizers: (c) and (nc), and (nc) has a lower energy $\Pi$ at convergence. At this stage, the standard Newton iteration still yields the global minimizer (nc). In Stage 3* when $\boldsymbol{u}_b$ surpasses a certain critical value $\boldsymbol{u}_b^c$, (c) has a lower energy than does (nc), yet the standard Newton iteration normally still converges to (nc). As such, a * is marked. In Stage 4, the load is so high that (c) is the only local (and global) minimizer and Newton iteration converges to (c), at a much higher load than the correct critical load $\boldsymbol{u}_b^c$. The crack is predicted to nucleate at stage 4 in the numerical simulation by the standard Newton method, much later than stage 3*, judging from the value of the energy functional $\Pi$. 

\begin{figure}[htbp]
\centering
	\includegraphics[width=6in,height=3in,keepaspectratio]{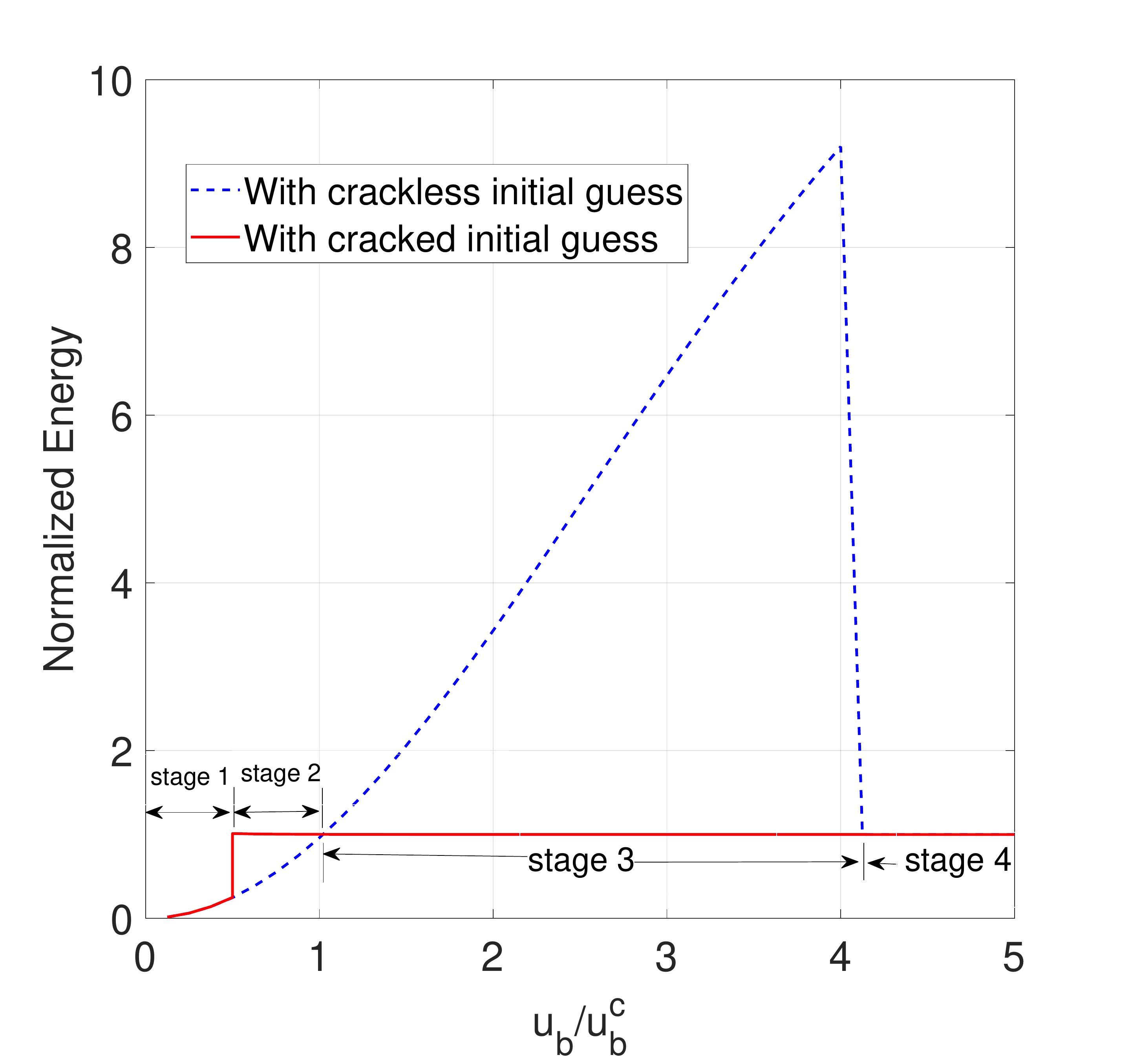}
	\caption{The total energies as  functions of the scaling factor of the displacement load, $u_{b}$, and the resulting four stages corresponding to Figure \ref{Diagram}. The correct global minimizer first follows the dashed line and upon $u_b/u_b^c\ge1$, follows the solid line. In contrast, the standard Newton method follows the dashed line all the way, and only when the load is much higher than $u_b^c$, a cracked solution is obtained.}
	\label{Solution}
\end{figure}

Another perspective of the issue is provided in Figure \ref{Solution} with the relevant energies of the converged results as a function of load $u_b$ with cracked and crackless initial guesses. Again, it can be seen that such difficulty gives rise to a much higher cracking load.

\subsection{Main idea of the algorithm}
\label{MainIdea}
The proposed scheme is based on the following considerations.
For the crack nucleation problem at hand, if a cracked initial guess is generated when such a cracked local minimizer is very likely to exist, simply following the Newton iteration scheme may give rise to a cracked solution at convergence, somewhat similar to the idea of numerical continuation methods \cite{allgower2003introduction}. Then a simple comparison of the two candidate solutions of their $\Pi$ values decides whether the cracked solution $\boldsymbol{u}_c$, or the crackless solution $\boldsymbol{u}_{nc}$, is more likely the global minimizer. 

\begin{figure}[htbp]
   \centering
   \includegraphics[width=12cm,keepaspectratio]{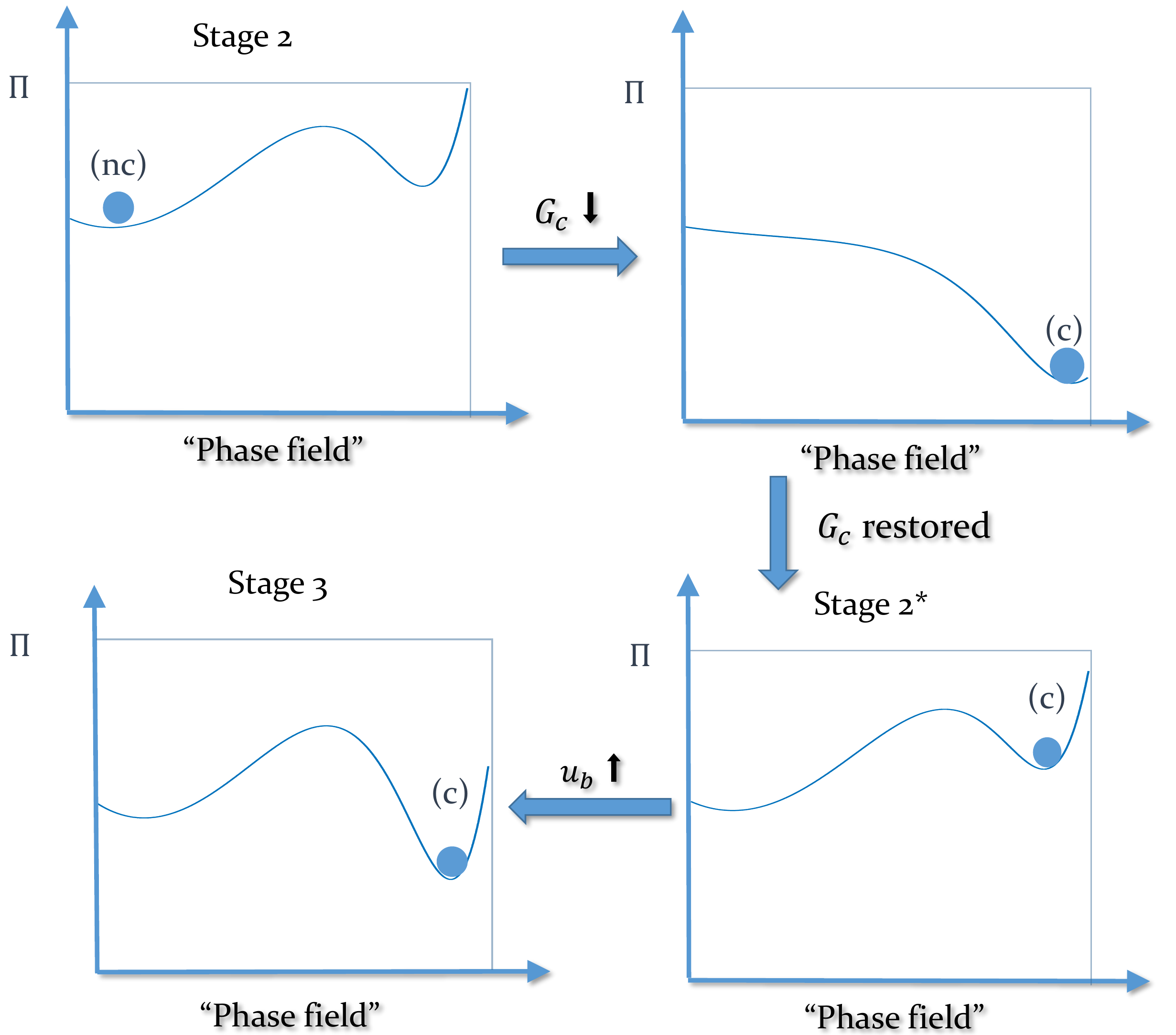}
   \caption{Illustration of the proposed algorithm. Assume Stage 2 (scaled from the counterpart in Figure \ref{Diagram} for clarity) is when \eqref{CrackCondition} is satisfied for the first time.}
    \label{Diagram2V2}
\end{figure}

We then progressively reduce $G_c$ to obtain a cracked initial guess (upper right subfigure) and then restore the value of $G_c$ to obtain a cracked candidate solution, Stage 2*. As the load $u_b$ further increases, the cracked candidate solution may eventually yield a lower energy $\Pi$, which will be accepted as the solution, as in Stage 3. The main idea is also illustrated in Figure \ref{Diagram2V2}.

Such a search does not need to be frequently performed. In fact, if the crackless solution $\boldsymbol{u}_{nc}$ is accepted for yielding a lower energy, at the next step, both the cracked and crackless solutions of the previous step will be used as the initial guesses for Newton iteration, as if they co-exist in different parallel universes, and hence the name ``parallel universe scheme.'' Of course if $\boldsymbol{u}_c$ is accepted instead, $\boldsymbol{u}_{nc}$ is discarded for future steps, per irreversibility.

\section{Method}
\label{sec:Method}
In Section \ref{BasicFormulations}, some basic formulations of the phase field approach to fracture and its discretization are introduced. In Section \ref{Newton iteration}, the staggered version of the standard Newton iteration is introduced. In Section \ref{TwoKeyIssue}, two necessary components of the proposed algorithm are introduced. In Section \ref{EntireAlgorithm}, the entire proposed algorithm is introduced. Readers interested in the entire algorithm can directly see Algorithm \ref{FlowChart1}.

\subsection{The phase field approach to fracture and its discretization}
\label{BasicFormulations}
We first specify some quantities for the functional \eqref{Energy equation}. Note that the formulas below are written for the plane strain case, and generalization to 3D is straightforward. First, we adopt the following form for the strain energy density $\psi$ proposed by Amor et al.~\cite{AMOR20091209}:
\begin{equation*}
\psi(\nabla\mathbf{u},d)=[(1-d)^2+k]\psi_{+}(\nabla\mathbf{u})+\psi_{-}(\nabla\mathbf{u}),
\end{equation*}
where $k$ is a smaller number usually taken as $10^{-10}$, and 
\begin{equation*}
\begin{aligned}
\psi_{+}=\frac{K}{2}\left(\frac{\trace\boldsymbol{\varepsilon}+\left|\trace\boldsymbol{\varepsilon}\right|}{2}\right)^{2}+\mu\left\|\deviatoric\boldsymbol{\varepsilon}\right\|^{2},\quad
\psi_{-}=\frac{K}{2}\left(\frac{\trace\boldsymbol{\varepsilon}-\left|\trace\boldsymbol{\varepsilon}\right|}{2}\right)^{2},
\end{aligned}
\end{equation*}
where $K>0$ is the bulk modulus and $\mu>0$ is the shear modulus, and the strain field is $\boldsymbol{\varepsilon}(\mathbf{u})=\left(\nabla\mathbf{u}+\nabla\mathbf{u}^{\mathrm{T}}\right)/2$. From now we adopt the Voigt notation to rewrite $\boldsymbol{\varepsilon}$ as a $3\times1$ vector in the Voigt notation. The trace of  $\boldsymbol{\varepsilon}$ is $\trace\boldsymbol{\varepsilon}=\boldsymbol{\varepsilon}\cdot\mathbf{1}$,  where $\mathbf{1}=[1,1,0]^T$ is the identity tensor expressed in the Voigt notation.
    The deviatoric part of $\boldsymbol{\varepsilon}$ is $\deviatoric\boldsymbol{\varepsilon}=\boldsymbol{\varepsilon}-(1/3)(\trace\boldsymbol{\varepsilon})\mathbf{1}$.

The domain $\Omega$ is discretized into a number of finite elements with $\boldsymbol{n}_{\rm{nodes}}$ nodes. The displacement and phase fieldsd are discretized accordingly as
\begin{equation*}
    \label{field discretization}
    \boldsymbol{u(x)}=\sum_{A=1}^{\boldsymbol{n}_{\rm{nodes}}}\mathbf{N}_{A}\mathbf{u}_{A}(\mathbf{x}),\quad d\mathbf{(x)}=\sum_{A=1}^{\boldsymbol{n}_{\rm{nodes}}}N_{A}d_{A}(\mathbf{x}),
\end{equation*}
where $N_{A}$ is the shape function of node $A$, and
\begin{equation*}
    \mathbf{N}_{A}=\begin{bmatrix} 
    N_{A}&0\\
    0&N_{A}\\
    \end{bmatrix}.
\end{equation*}
In the sequel, $\mathbf{u}$ and $\mathbf{d}$ denote the nodal displacement and phase field values, respectively.

A necessary condition of the minimization of \eqref{Energy equation} is  $\mathbf{R}^\mathbf{u}=\partial\Pi/\partial\mathbf{u}=\mathbf{0}$ and $\mathbf{R}^\mathbf{d}=\partial\Pi/\partial\mathbf{d}=\mathbf{0}$. Vectors $\mathbf{R}^\mathbf{u}$ and $\mathbf{R}^\mathbf{d}$ are also called residual vectors.
The explicit expressions of the residual vectors with respect to node $A$ are:
\begin{align*}
        \mathbf{R}^{\mathbf{u}}_{A}(\mathbf{u},\mathbf{d})&=\int_{\Omega} (\mathbf{B}_{A}^{\mathbf{u}})^{T}\boldsymbol{\sigma}\mathrm{d}\Omega,\\
        \mathbf{R}^{d}_A(\mathbf{u},\mathbf{d}) &=\int_{\Omega} \left[2\left(1-d\right)\psi_{+}N_{A}+G_{c}\left(\frac{ N_{A}d}{\ell}+\ell (\mathbf{B}_{A}^{d})^{T}\cdot\nabla d\right)\right]\mathrm{d}\Omega,
\end{align*}
         where 
    \begin{equation*}
    \boldsymbol{\sigma}=[(1-d)^{2}+k]\left(\frac{K}{2}(\trace{\boldsymbol{\varepsilon}}+\left|\trace\boldsymbol{\varepsilon}\right|)\mathbf{1}+2\mu\deviatoric{\boldsymbol{\varepsilon}}\right)+\frac{K}{2}(\trace\boldsymbol{\varepsilon}-\left|\trace\boldsymbol{\varepsilon}\right|)\mathbf{1},
    \end{equation*}
    $\mathbf{B}_{A}^{\mathbf{u}}$ is the strain-displacement matrix block  of node $A$:
    \begin{equation*}
        \mathbf{B}_{A}^{\mathbf{u}}=\begin{bmatrix}
        N_{A,x} &0 \\
        0 & N_{A,y} \\
         N_{A,y}& N_{A,x}
        \end{bmatrix},
    \end{equation*}
and $\mathbf{B}_{A}^{d}$ is defined as: 
    \begin{equation*}
 \mathbf{B}_{A}^{d}=\begin{bmatrix}
 N_{A,x}&N_{A,y}
 \end{bmatrix}^{T}.
    \end{equation*}

        The tangent stiffness vectors are the partial derivatives of the residuals with respect to the degrees of freedom. In a staggered algorithm, only entries relating two like degrees of freedom (both displacement or both phase field) are needed. In particular, the tangent stiffness entry related to node $A$ and node $B$ take the following forms 
          \begin{align*}
        \mathbf{K}^{\mathbf{u}}_{AB}(\mathbf{u},\mathbf{d})&=\int_\Omega(\mathbf{B}_{A}^{\mathbf{u}})^{T}\mathbb{C}[\boldsymbol{\varepsilon}(\mathbf{u}),d]\mathbf{B}_{B}^{\mathbf{u}}\;\mathrm{d}\Omega,\\
        \mathbf{K}^{d}_{AB}(\mathbf{u},\mathbf{d})&=\int_\Omega\left[\left(2\psi_{+}+\frac{G_{c}}{\ell}\right)N_A N_B+G_{c}\ell(\mathbf{B}_{A}^{d})^{T}\mathbf{B}_{B}^{d}\right]\mathrm{d}\Omega,
    \end{align*}
    where  $\mathbb{C}[\boldsymbol{\varepsilon}(\mathbf{u}),d]$, a $3\times3$ matrix, is given by
    \begin{equation*}
    \begin{aligned}
     \mathbb{C}[\boldsymbol{\varepsilon}(\mathbf{u}),d] =
     \left[(1-d)^2+k\right]\left[KH(\trace\boldsymbol{\varepsilon})\mathbf{1}\mathbf{1}^{T}+2\mu\left(\mathbb{I}-\frac{\mathrm{1}}{\mathrm{3}}\mathbf{1} \mathbf{1}^{T} \right)\right]
     +KH(-\trace{\boldsymbol{\varepsilon}})\mathbf{1} \mathbf{1}^{T},
     \end{aligned}
    \end{equation*}
    where $H(\cdot)$ is the Heaviside function and $\mathbb{I}$ is the $3\times3$ identity matrix.

 \subsection{The staggered version of the standard Newton iteration}
 \label{Newton iteration}
We next introduce the standard staggered Newton iteration, upon which the proposed algorithm is based. The Newton scheme attempts to find the solution of nonlinear equations by iteration. In the present problem, let $\mathbf{u}^{(m)}$ and $\mathbf{d}^{(m)}$ denote the nodal displacement and the phase field of the $m$th iteration, respectively. At a certain load, starting from initial guesses $(\mathbf{u}^{(0)}, \mathbf{d}^{(0)})$, normally the solution of the previous load step, the Newton iteration  solves $\mathbf{u}^{(m+1)}$ or $\mathbf{d}^{(p+1)}$ from $\mathbf{u}^{(m)}$ and $\mathbf{d}^{(p)}$, $m,p=1,2,\ldots$, using the following equations
\begin{equation*}
\label{Solve_u}
\mathbf{K_{u}}\left(\mathbf{u}^{(m)},\mathbf{d}^{(p)}\right)\Delta\mathbf{u}=-\mathbf{R_{u}}\left(\mathbf{u}^{(m)},\mathbf{d}^{(p)}\right),\quad
\mathbf{u}^{(m+1)}=\mathbf{u}^{(m)}+\Delta\mathbf{u},
\end{equation*} 
and
\begin{equation*}
\label{Solve_d}
\mathbf{K}_{d}\left(\mathbf{u}^{(m)},\mathbf{d}^{(p)}\right)\Delta \mathbf{d}=-\mathbf{R}_{d}\left(\mathbf{u}^{(m)}, \mathbf{d}^{(p)}\right),\quad
\mathbf{d}^{(p+1)}=\mathbf{d}^{(p)}+\Delta \mathbf{d}.
\end{equation*}


As mentioned before, this standard Newton iteration scheme is known to highly overestimate the cracking load for crack nucleation problems. The key is to find cracked initial guesses when the criterion \eqref{CrackCondition} is met, and perform Newton iterations based on both cracked and crackless initial guesses, and hence the name the parallel universe scheme.
 \subsection{Details of the proposed optimization algorithm}
 \label{TwoKeyIssue}
As mentioned in Section \ref{MainIdea}, the key issue boils down to two questions: What criterion should trigger the search for a cracked initial guess, and how to find it. 
\subsubsection{Criterion for triggering the scheme}
For the first question, a very useful criterion turns out to be that the maximum principal stress, $\sigma_{\max}$, at any point exceeds a certain value $\sigma_v$, termed the stress of vigilation. Inspired by \cite[Equations (5) and (6)]{TANNE201880}, as we are using the AT2 model, we define
\begin{equation*}
\label{sigma_c}
\sigma_{c}=\sqrt{\frac{27G_{\rm{c}}E}{256\ell(1-\nu^{2})}},
\end{equation*}
where $E$ and $\nu$ are Young's modulus and Poisson's ratio, respectively. Note that if $\ell$ is treated as a material parameter as in \cite{TANNE201880}, then $\sigma_{c}$ coincides with the tensile strength of the material. 
The criterion is then given by
\begin{equation}
\label{CrackCondition}
\sigma_{\max}\geq\sigma_v=\frac{\sigma_{c}}{\alpha},
\end{equation}
where $\alpha$ is the safety factor.
For convenience, we define the first load when \eqref{CrackCondition} is met as $u_b^v$. Note that \eqref{CrackCondition} merely triggers the \emph{search} for a converged cracked solution; whether to accept such a solution depends on its competition with the crackless candidate. Empirically a value of $\alpha=1$ is sufficiently large to ensure $u_b^v<u_b^c$, i.e., to avoid missing the critical load $u_b^c$. 
More details are in Section \ref{sec:Disccusion}.


\subsubsection{Obtaining the cracked initial guess}
\begin{figure}[htbp]
    \centering
    \includegraphics[width=12cm]{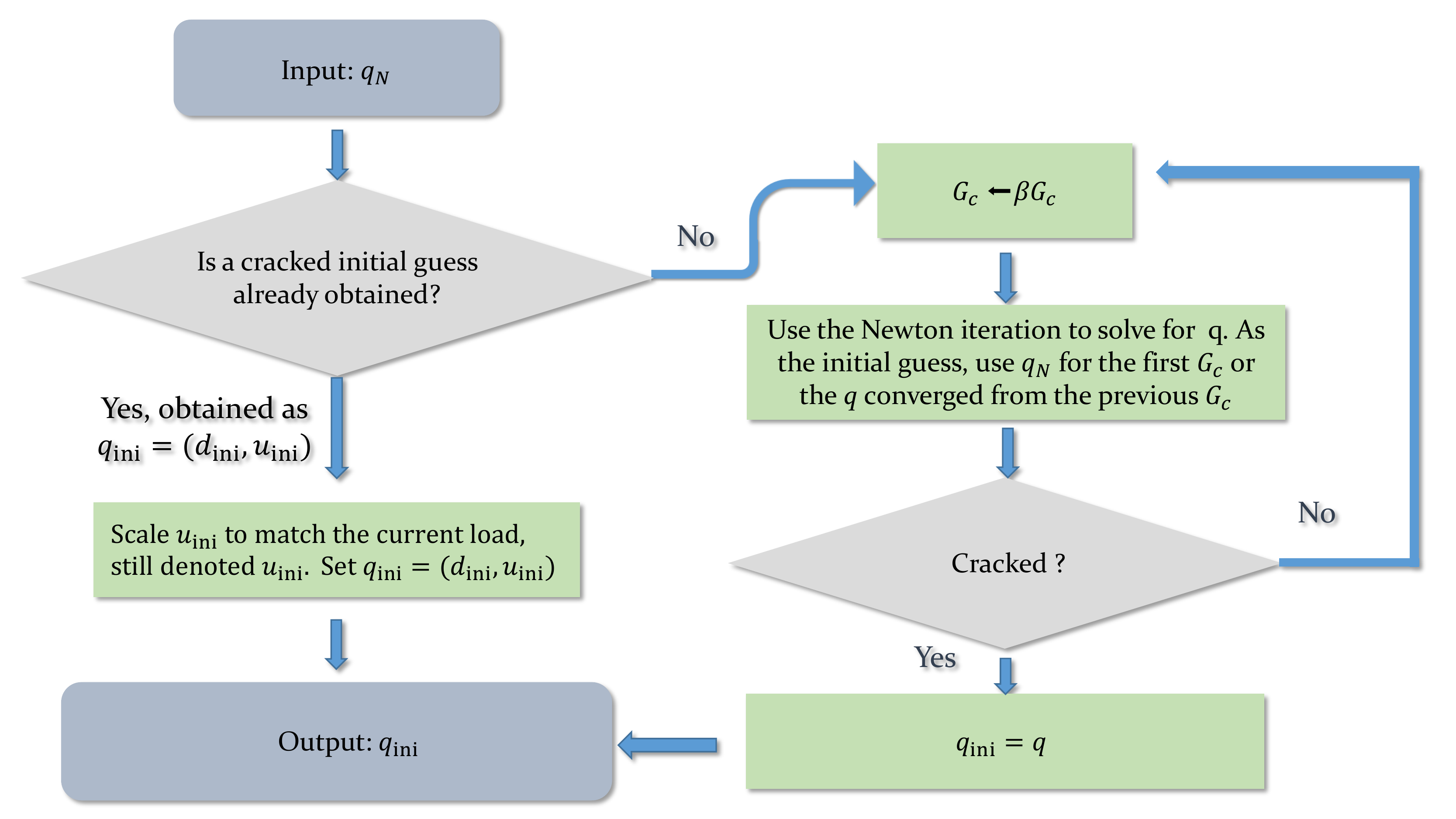}
    \caption{Flowchart showing the algorithm to obtain a cracked initial guess.}
    \label{FlowChart2}
\end{figure}

For the second question, in order to find a cracked initial guess when \eqref{CrackCondition} is met, we progressively reduce $G_c$ until the converged solution has one or more cracks, with an inspiration from the accelerated sampling schemes for molecular dynamics \cite{Laio12562}. In the implementation, this criterion is written as $\max d(\Omega)\geq 0.9$. The principle is illustrated in Figure \ref{Diagram2V2} and the progress is shown in Figure \ref{FlowChart2}. Here $q_N=(d_N,\boldsymbol{u}_N)$ denotes the converged phase field and displacement field of the $N$th load step, and $q^{NC}$ and $q^{C}$, sometimes with a subscript, denote a crackless candidate solution and a cracked candidate solution, respectively. Let us reiterate that the computation shown in Figure \ref{FlowChart2} is expensive, but only needs to be carried out only once, and only after \eqref{CrackCondition} is met.

\subsection{The entire proposed algorithm}
\label{EntireAlgorithm}
\begin{figure}[htbp]
    \centering
    \includegraphics[width=12cm]{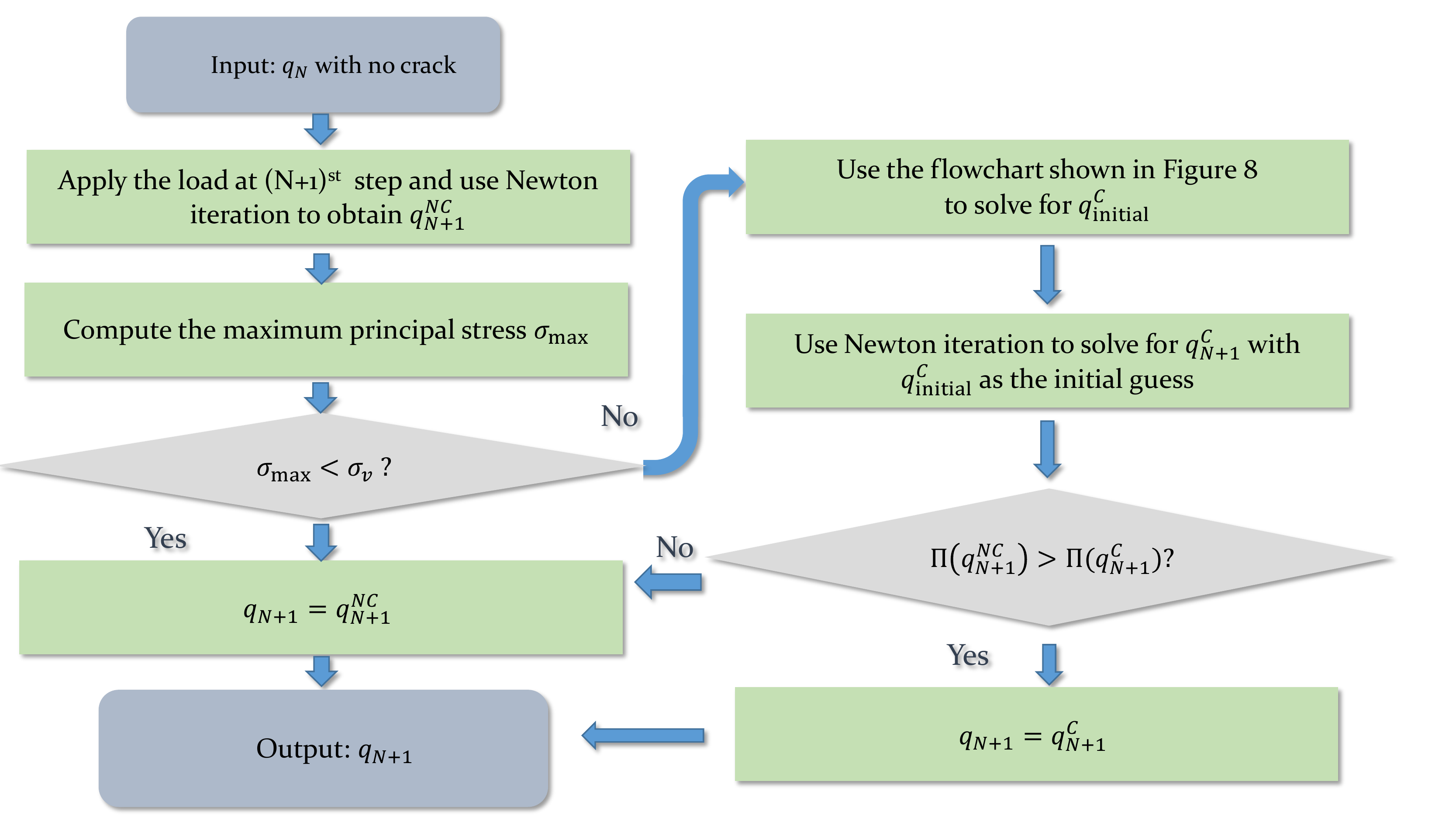}
    \caption{Flowchart representing the proposed load-stepping algorithm from step $N$ to step $N+1$ for the stage when there is no crack in the solid at $N$. Here $q_N=(d_N, \boldsymbol{u}_N)$.}
    \label{FlowChart1}
\end{figure}

The proposed algorithm is described in Figure \ref{FlowChart1}.  

\section{Numerical examples}
\label{sec:Numerical}

We showcase the proposed algorithm  with three examples. For all of them triangular elements and standard first-order finite element shape functions are employed.

\label{1d2dProblem}
\subsection*{Example 1: Tensile experiment on a fiber-reinforced matrix}
\begin{figure}[htbp]
    \centering
    \subfloat[]{
\includegraphics[width=4in,height=2in,keepaspectratio]{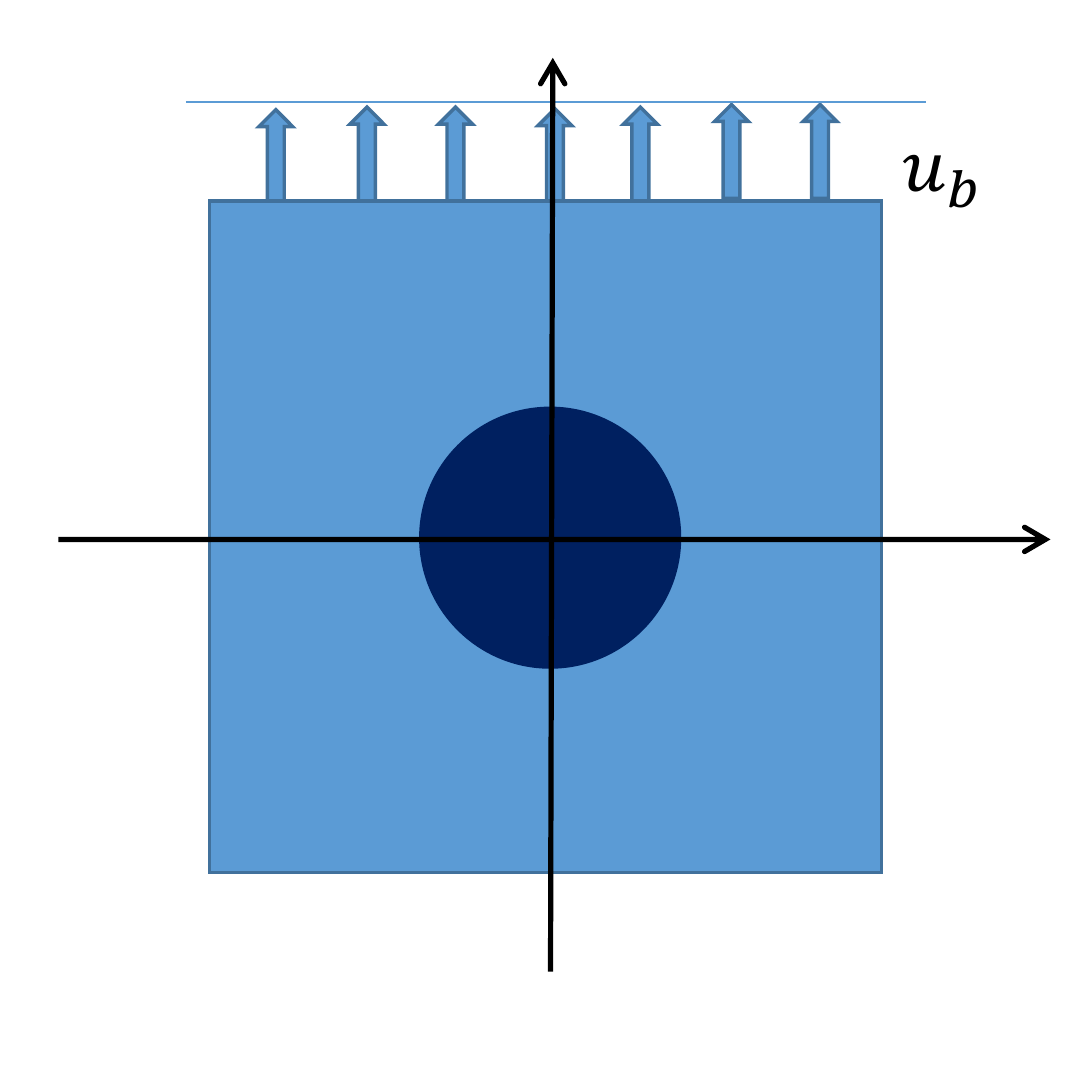}
\label{2dFigure3}}
\quad
\subfloat[]{
\includegraphics[height=2in,keepaspectratio]{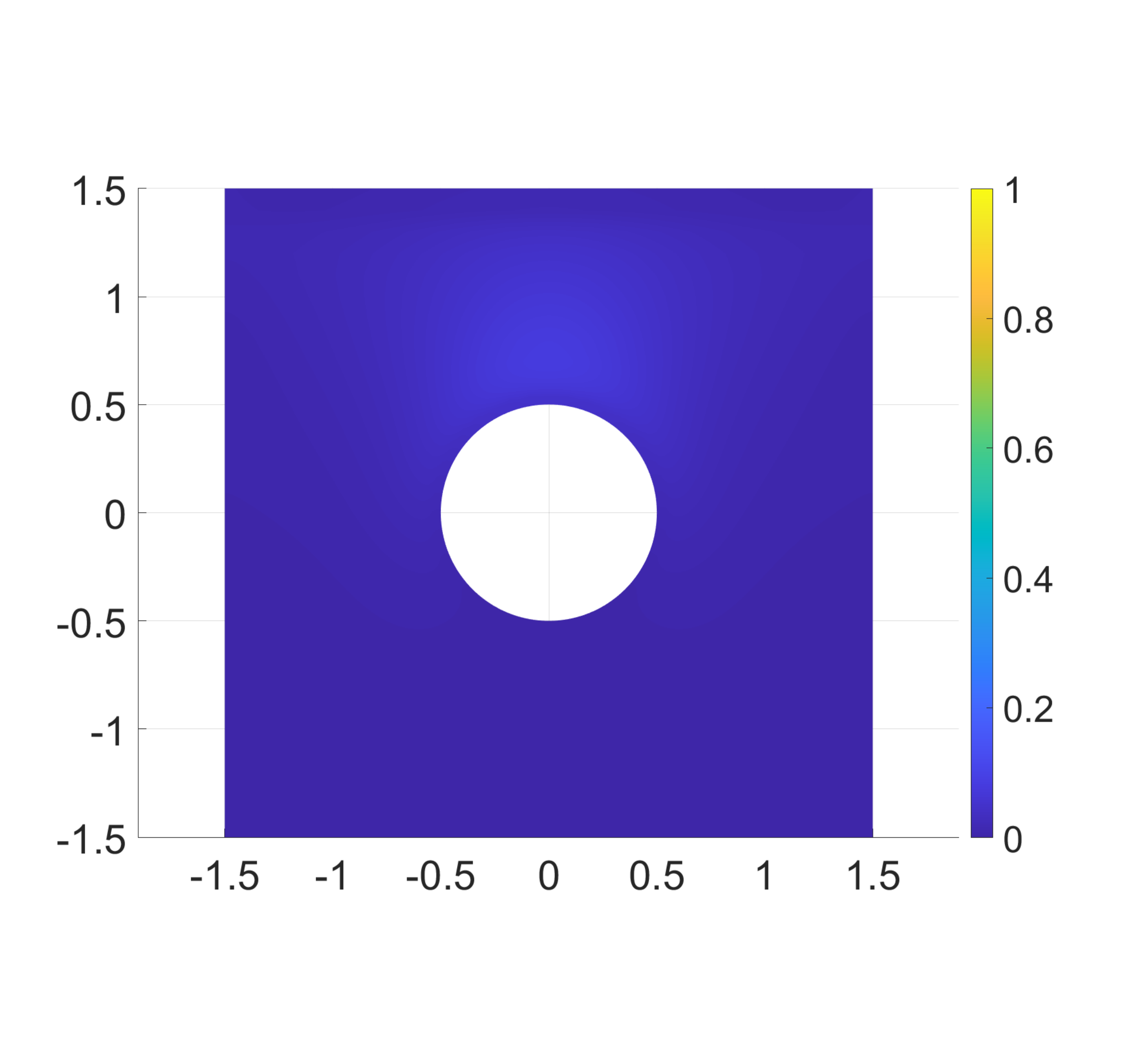}
\label{N1stressConcentration}
}
\quad 
\subfloat[]{
\includegraphics[width=4in,height=2in,keepaspectratio]{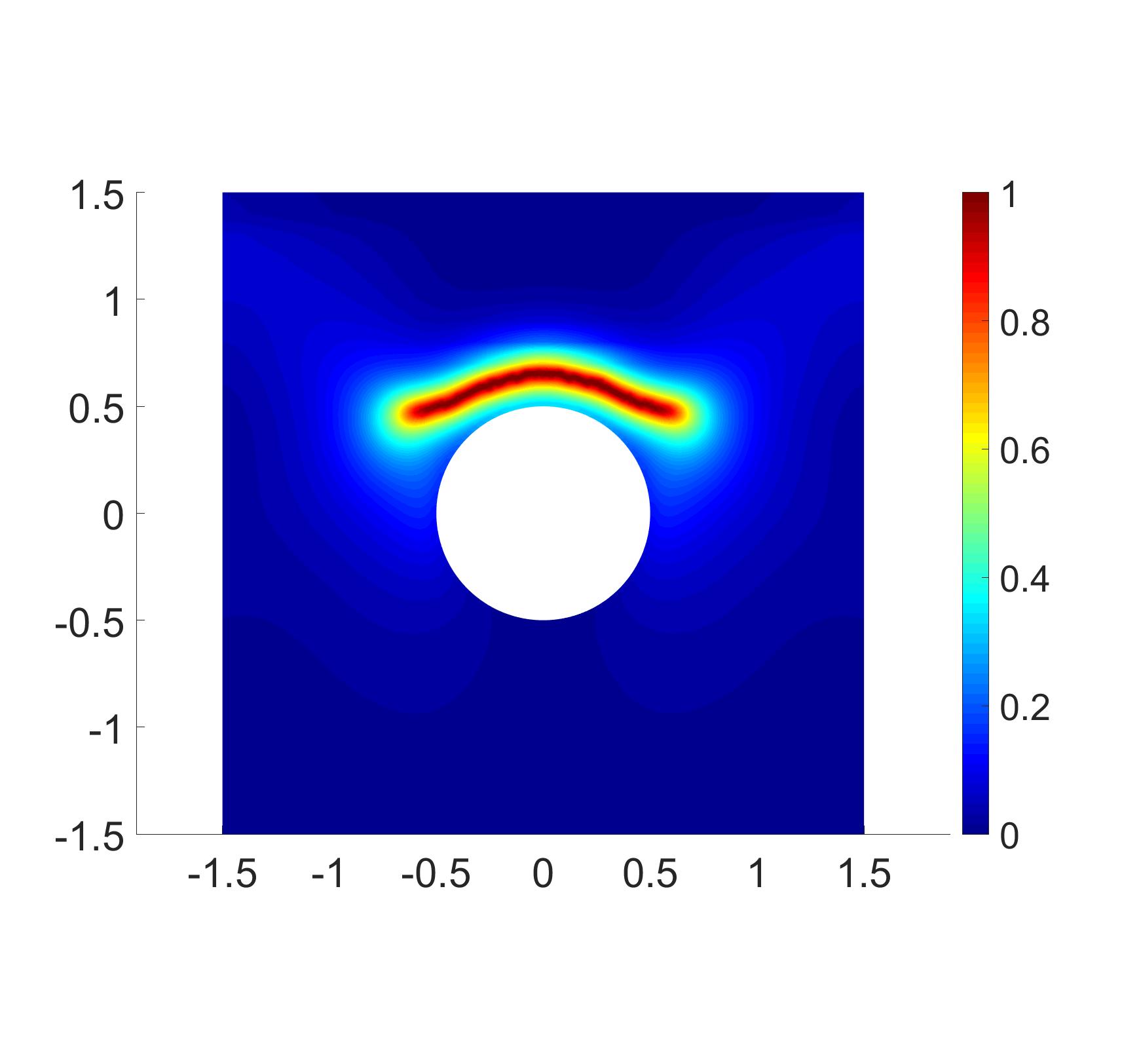}
\label{Numerical3PlotInitialGuess}}
\quad 
\subfloat[]{
\includegraphics[width=4in,height=2in,keepaspectratio]{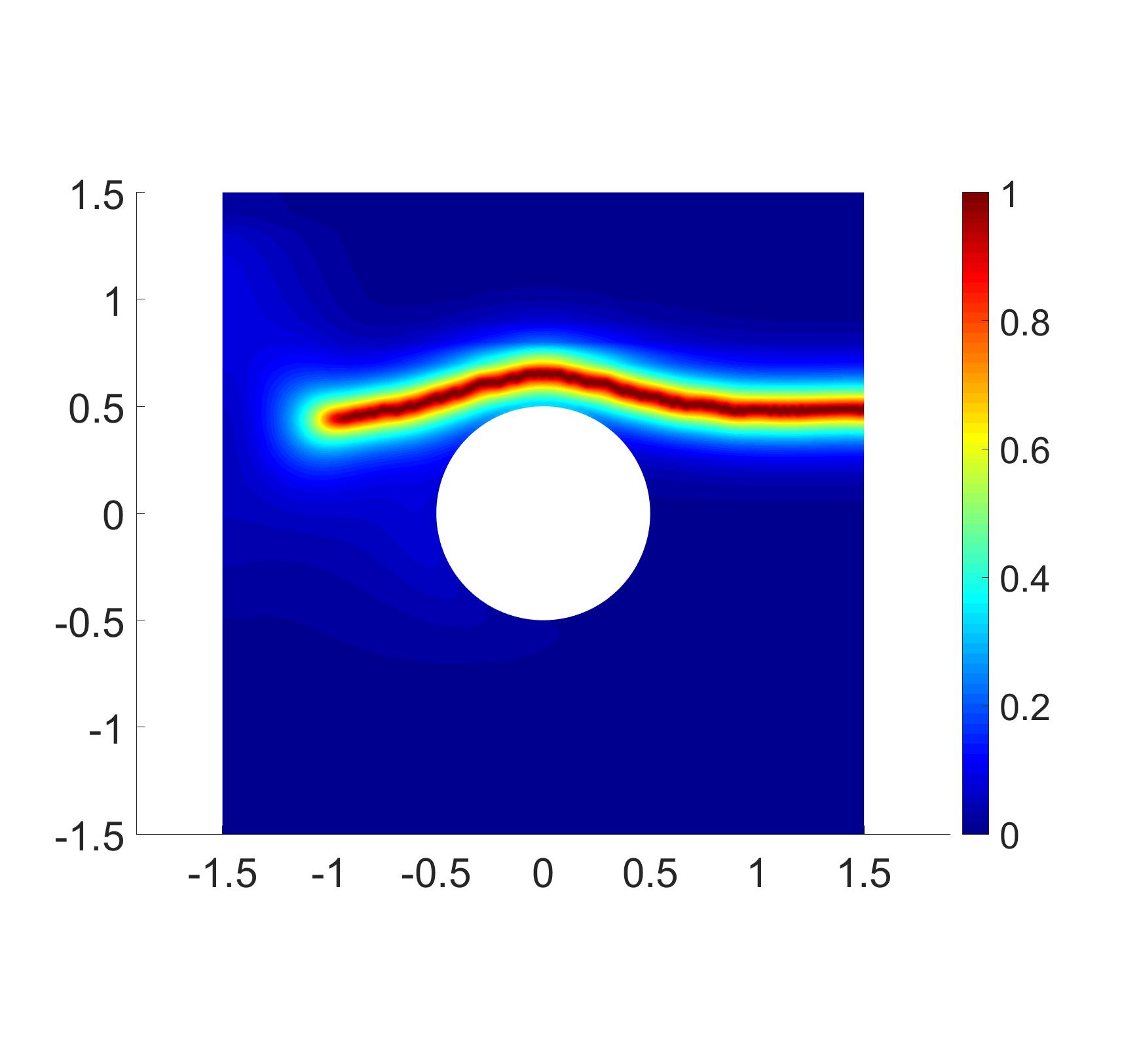}
\label{Numerical3Plot1}}
\quad 
\subfloat[]{
\includegraphics[width=4in,height=2in,keepaspectratio]{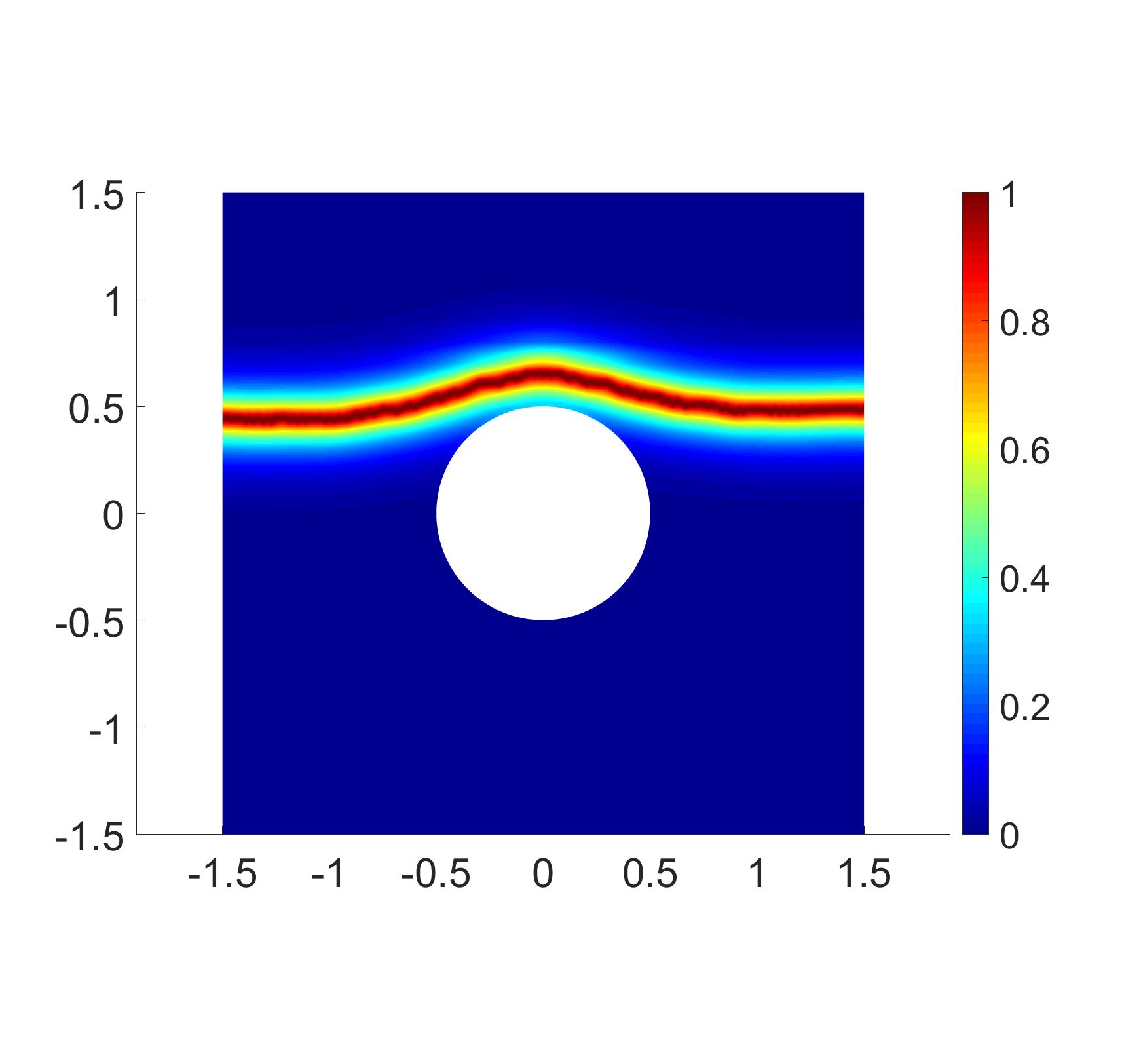}
\label{Numerical3Plot2}}
\caption{Crack nucleation example of the tensile experiment on a fiber-reinforced composite. (a) Schematic. (b)Phase field at the load of vigilance $u_b=u_b^v=0.125$. (c) Phase field of the cracked initial guess obtained at the load of vigilance $u_b=u_b^v=0.125$. Note that upon convergence, the energy with this cracked candidate is higher than th at with the crackless candidate when $u_b<u_b^c=0.260$; when $u_b>u_b^c$, the energy relations are reversed, and then the cracked candidate is accepted, i.e., crack nucleation is proclaimed. (d) Phase field of the domain when $u_b=0.410$. The crack asymmetrically propagates and reaches the right side of the domain. (e) Phase field of the domain when $u_b=0.450$. The crack then propagates and reaches the left side of the domain.}
\label{fig:Example1}
\end{figure}

We first verify the proposed algorithm with the simulation of a tensile experiment on a fiber-reinforced matrix by Bourdin et al.~\cite{BOURDIN2000797}. Consider a composite initially occupying the square $(-L/2,L/2)^2$ with some $L>0$, and the crackless solid matrix initially occupying the domain $\Omega$, where \[\Omega=\left\{(x,y)\in\mathbb{R}^2:-\frac{L}{2}<x<\frac{L}{2}, -\frac{L}{2}<y<\frac{L}{2}, x^2+y^2>R^2 \right\},\] with $0<R<L$, see Figure \ref{2dFigure3}. The fiber is assumed rigid and fixed with an external device, i.e., the boundary conditions on the interface 
$\left\{x^2+y^2=R^2 \right\}$,
is $(u_{x},u_{y})=(0,0)$.
The boundary conditions on the upper edge $(-L/2,L/2)\times\{ L/2\}$ is $(u_x,u_y)=(0,u_b)$, where $u_{b}$ increases from zero quasistatically until the solid is completely fractured. The left, right and lower edges $\{\pm L/2\}\times(-L/2,L/2)\cup(-L/2,L/2)\times\{ -L/2\}$ are traction free. 
The parameters for the simulation are listed in Table \ref{Tab:parameters1}.

\begin{table}[htbp]
	\centering
	\caption{Material parameters for Example 1}
	\begin{tabular}{lccc}
		\toprule
		Parameter & Symbol & Value (in non-dimensionalized units) \\
		\hline
		Young's modulus & $E$ &4000  \\
		Poisson's ratio & $\nu$ & 0.2  \\
		Critical energy release rate & $G_c$ & 100  \\
		Phase field length scale parameter & $\ell$ & 0.1  \\
			Side length of the domain & $L$ & 3 \\
			Radius of the circle in the center& $R$ & 0.5 \\
		\bottomrule
	\end{tabular}
\label{Tab:parameters1}
\end{table}

For this example, the load of vigilance is found to be $u_b^v=0.125$. At this load, the cracked initial guess is obtained and is plotted in Figure \ref{Numerical3PlotInitialGuess}. At this load, two parallel universes are initiated, yet until $u_b<u_b^c=0.260$, the crackless candidate still gives a lower energy and is thus accepted as the minimizer. In contrast, when $u_b>u_b^c$, the cracked candidate is accepted and this is the crack nucleation load predicted by the model, and the crackless candidate is discarded. Then the crack is observed to propagate first from the right tip and then from the left tip, reaching the left and right boundaries of the domain when $u_b=0.410$ and $u_b=0.450$, respectively. The results are almost the same as those in \cite[Chapter 4.1]{BOURDIN2007411} using backtracking.

\subsection*{Example 2: Square domain with a hole at the center}
\label{Verification}
\begin{figure}
    \centering
    \subfloat[]{
\includegraphics[height=1.7in,keepaspectratio]{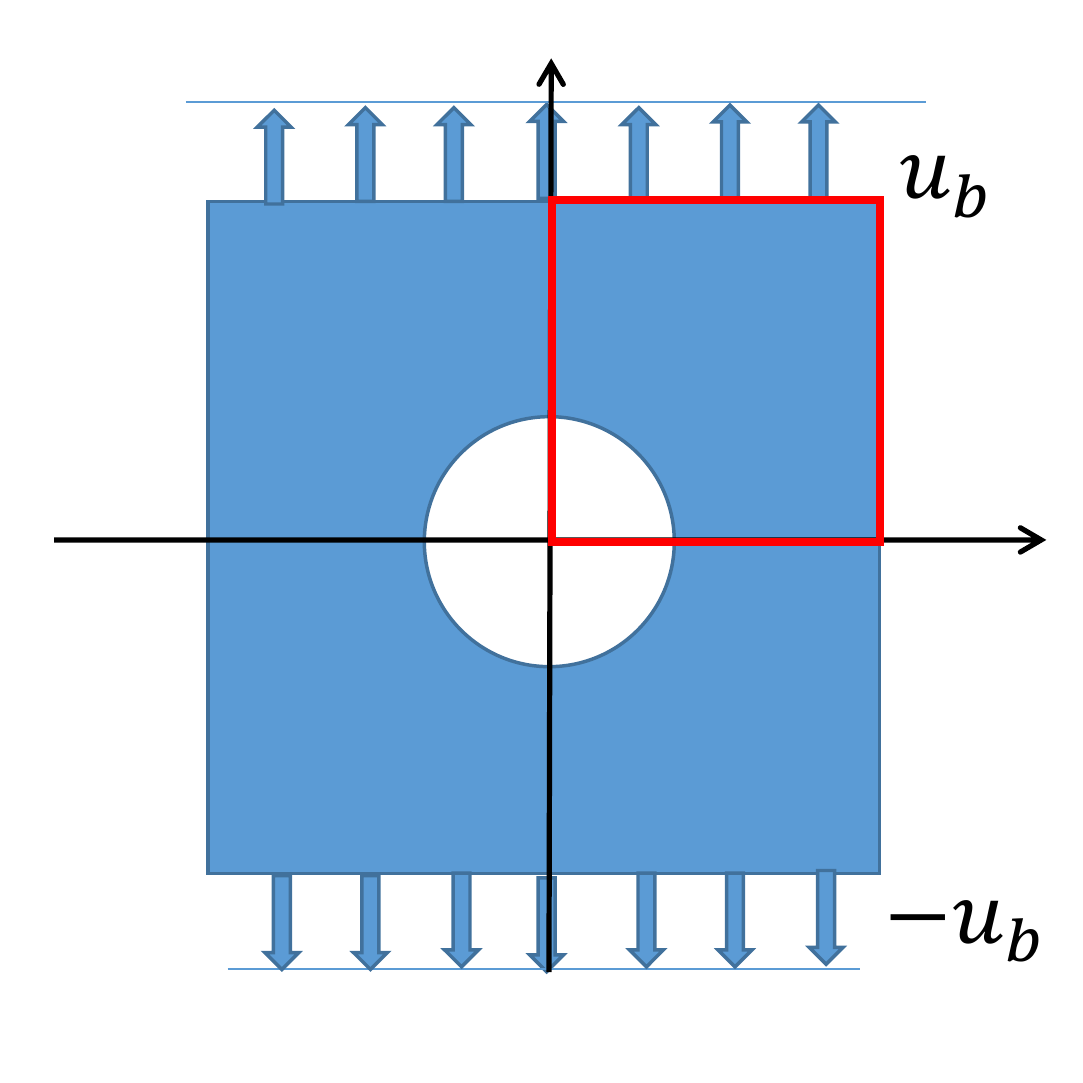}
\label{2dFigure1}}
\quad
\subfloat[]{
\includegraphics[height=1.7in,keepaspectratio]{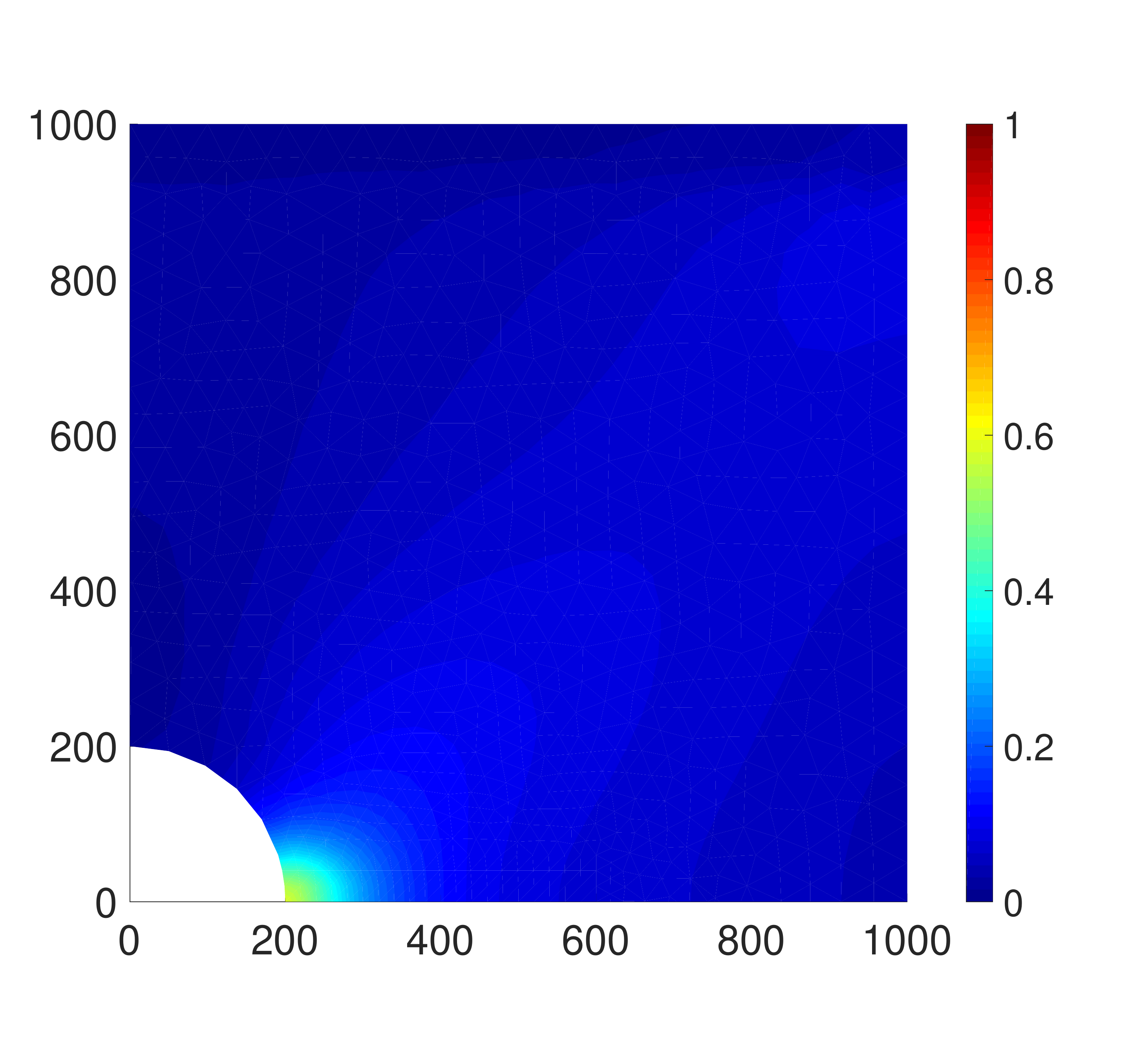}
\label{PhasefieldResultComparisonResult1}
}
\quad
    \subfloat[]{
\includegraphics[height=1.7in,keepaspectratio]{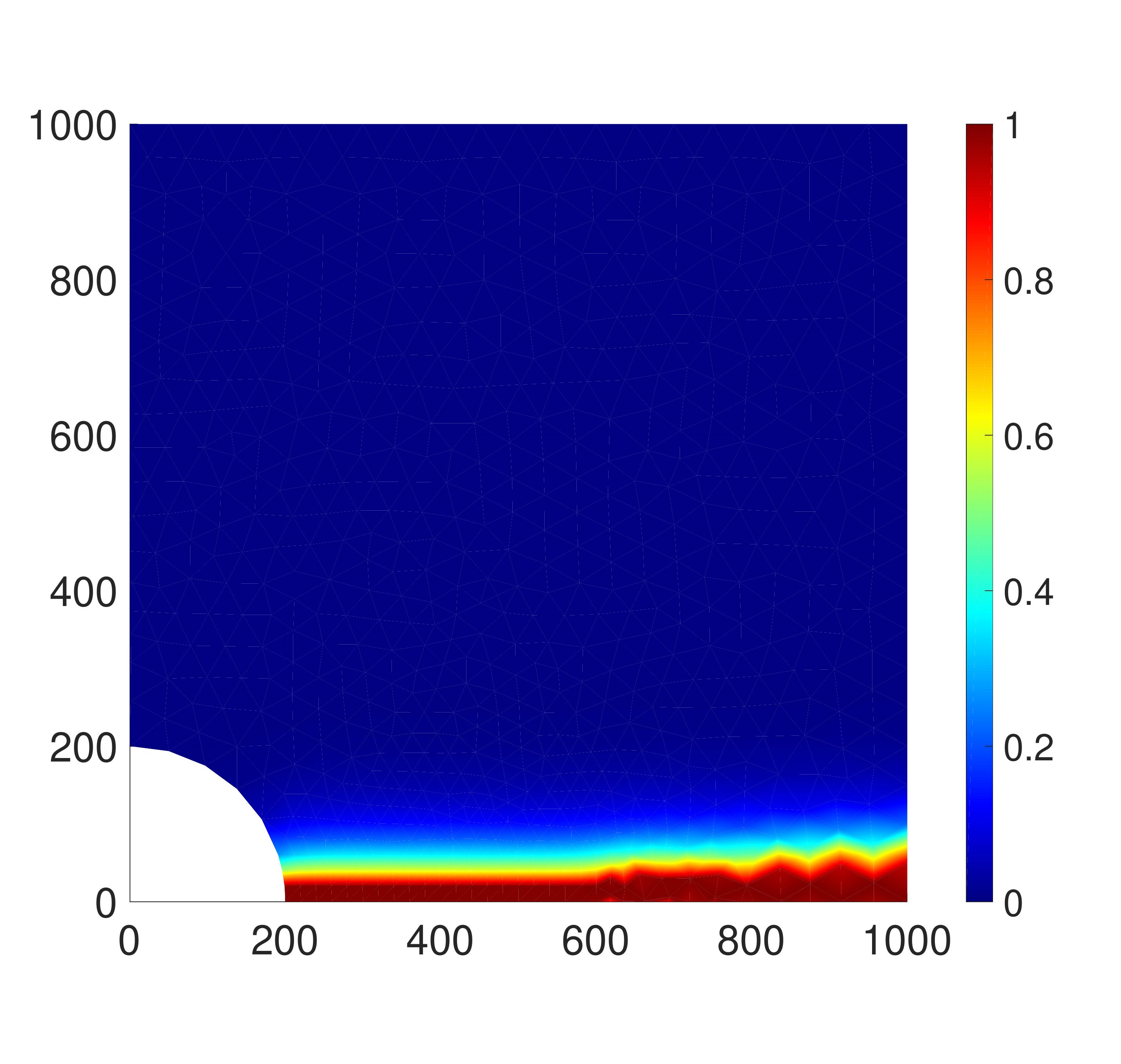}
\label{PhasefieldResultComparisonIntialGuess}
}
\quad
\subfloat[]{
\includegraphics[height=1.7in,keepaspectratio]{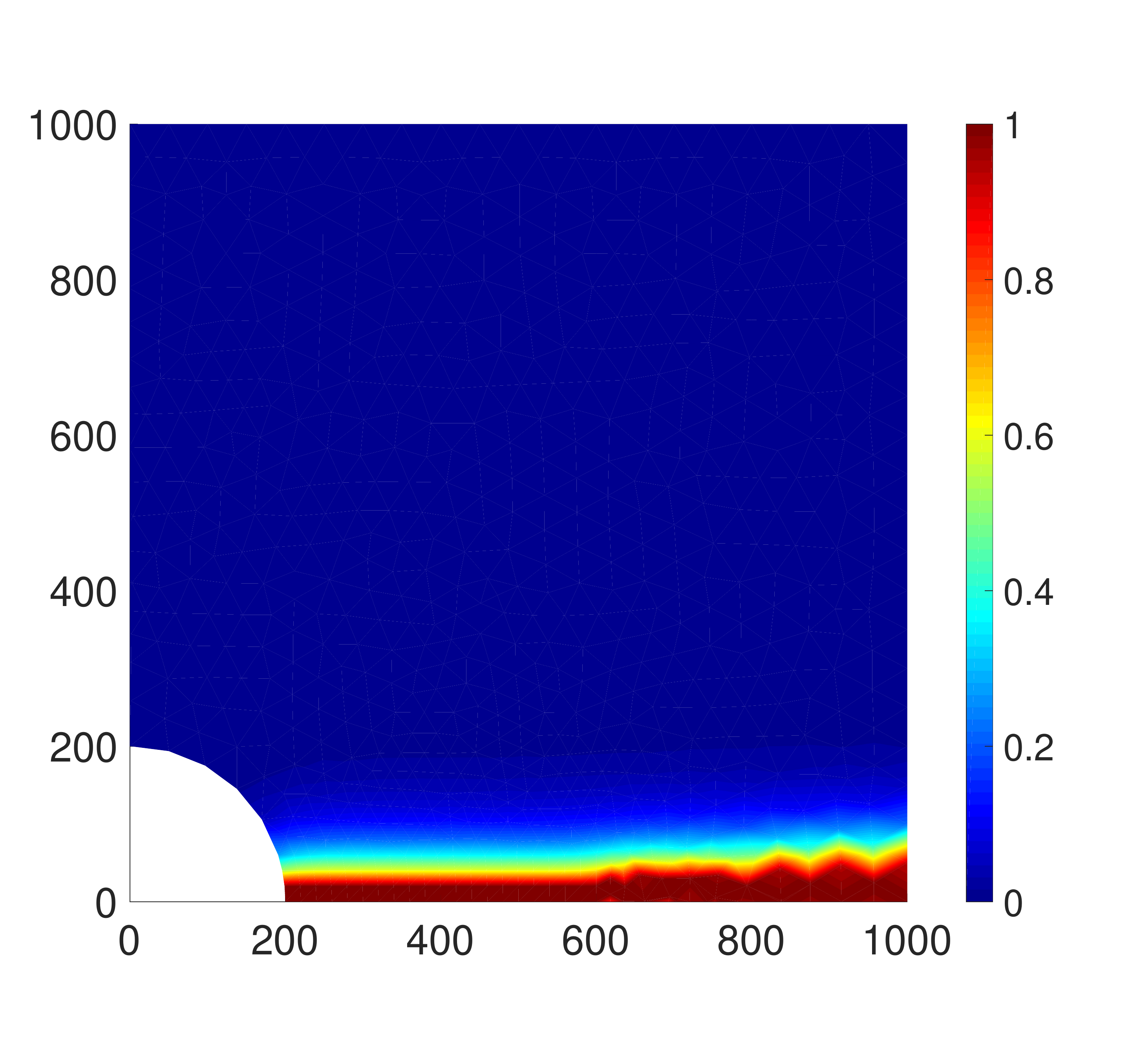}
\label{PhasefieldResultComparisonResult2}
}
\quad

    \caption{Crack nucleation example of the holed square. (a) Schematic. (b) Phase field in the red box in (a) when $u_b=u_b^v=0.092$mm. (c) Phase field of the cracked initial guess of the red block in (a) obtained when $u_b=u_b^v=0.092$mm. The result with this cracked initial guess at this time is almost the same as it. (d) Phase field  of the red block in (a) when $u_b=u^c_b=0.175$mm with the proposed algorithm, which is almost the same as (c). Compared with \cite[Figure 15]{TANNE201880}, the proposed algorithm has the ability to achieve the same converged results before and after the nucleation of the crack.}
    \label{PhasefieldResultComparison}
\end{figure}

We then verify the proposed algorithm with a crack initiation problem of a square domain with a hole under plane-strain loading, as studied by Tann\'{e} et al.~\cite{TANNE201880}. Consider a crackless solid initially occupying the domain $\Omega$, where \[\Omega=\left\{(x,y):-\frac{L}{2}<x<\frac{L}{2}, -\frac{L}{2}<y<\frac{L}{2}, x^2+y^2>R^2 \right\},\] with $L>R>0$, see Figure \ref{2dFigure1}. The boundary conditions on the upper and lower edges $(-L/2,L/2)\times\{\pm L/2\}$ are $(u_x,u_y)=(0,\pm u_b)$, where $u_{b}$ increases from zero quasistatically until the solid is completely fractured. The left and right edges $\{\pm L/2\}\times(-L/2,L/2)$ and also the hole surface $\{x^2+y^2=R^2\}$ are traction free. 
The parameters for the simulation are listed in Table \ref{Tab:parameters2}.

\begin{table}[htbp]
	\centering
	\caption{Material parameters for problem 2}
	\begin{tabular}{lccc}
		\toprule
		Parameter & Symbol & Value &Unit \\
		\hline
		Young's modulus & $E$ &210 &GPa \\
		Poisson's ratio & $\nu$ & 0.3 &-- \\
		Critical energy release rate & $G_c$ & 6750 & N/m \\
		Phase field length scale parameter & $\ell$ & 40 & mm \\
			Side length of the domain & $L$ & 2000 &mm\\
			Radius of the circle in the center& $R$ & 200 &mm\\
		\bottomrule
	\end{tabular}
\label{Tab:parameters2}
\end{table}



As the displacement load $u_b$ increases, the phase field rises at the boundary of the hole as a result of stress concentration, see Figure \ref{PhasefieldResultComparisonResult1}. At the load of vigilance $u_b=u_b^v=0.092$mm, a cracked initial guess is obtained and its converged phase field is plotted in Figure \ref{PhasefieldResultComparisonIntialGuess}. As before, at this load the cracked candidate still has a higher energy than the crackless counterpart, and this is when ``parallel universes'' are initiated. When $u_b$ reaches $u^c_b=0.175$mm, from which point crack nucleation is declared and only the cracked candidate is accepted. Overall speaking, the proposed algorithm obtains the same result in each instant as in \cite{TANNE201880}.

We also carried out the same example with the standard Newton iteration and the backtracking method \cite{BOURDIN2007411}. With the standard Newton iteration, the crack nucleates when $u_b=0.202$mm, which is clearly much higher than $u_b^c$.

The backtracking algorithm retraces from $u_b=0.202$mm to $u_b=0.175$mm and then converges to a cracked result, which effectively yields the same critical load for cracking as the proposed algorithm does. 

In summary, the proposed algorithm is equally accurate in terms of prediction for $u_b^c$, the critical load for cracking, and both methods are superior to the standard Newton iteration. In terms of efficiency in retrospect, the backtracking method needs to calculate for a load 15\% higher than $u_b^c$ for retracing, while the proposed method ``gets prepared'' at 53\% of $u_b^c$ and doubles the solution efforts until $u_b^c$ is reached. A quantitative efficiency comparison is offered for the next example.

\subsection*{Example 3: A homogeneous square domain}
\label{Comparison to Backtracking algorithm}

\begin{figure}
    \centering
    \subfloat[]{
\includegraphics[width=4in,height=2in,keepaspectratio]{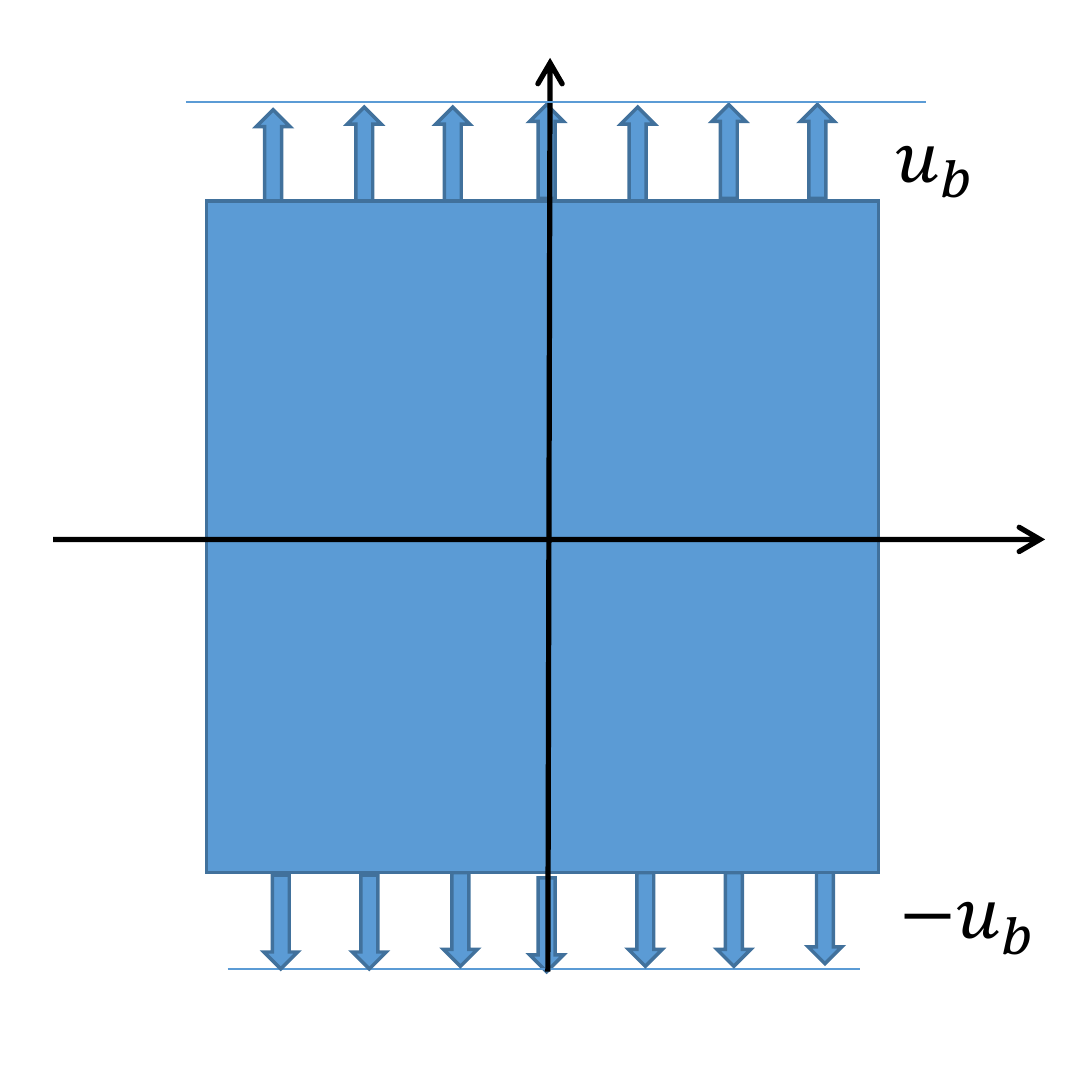}
\label{2dFigure2}}
\quad
\subfloat[]{
\includegraphics[width=4in,height=2in,keepaspectratio]{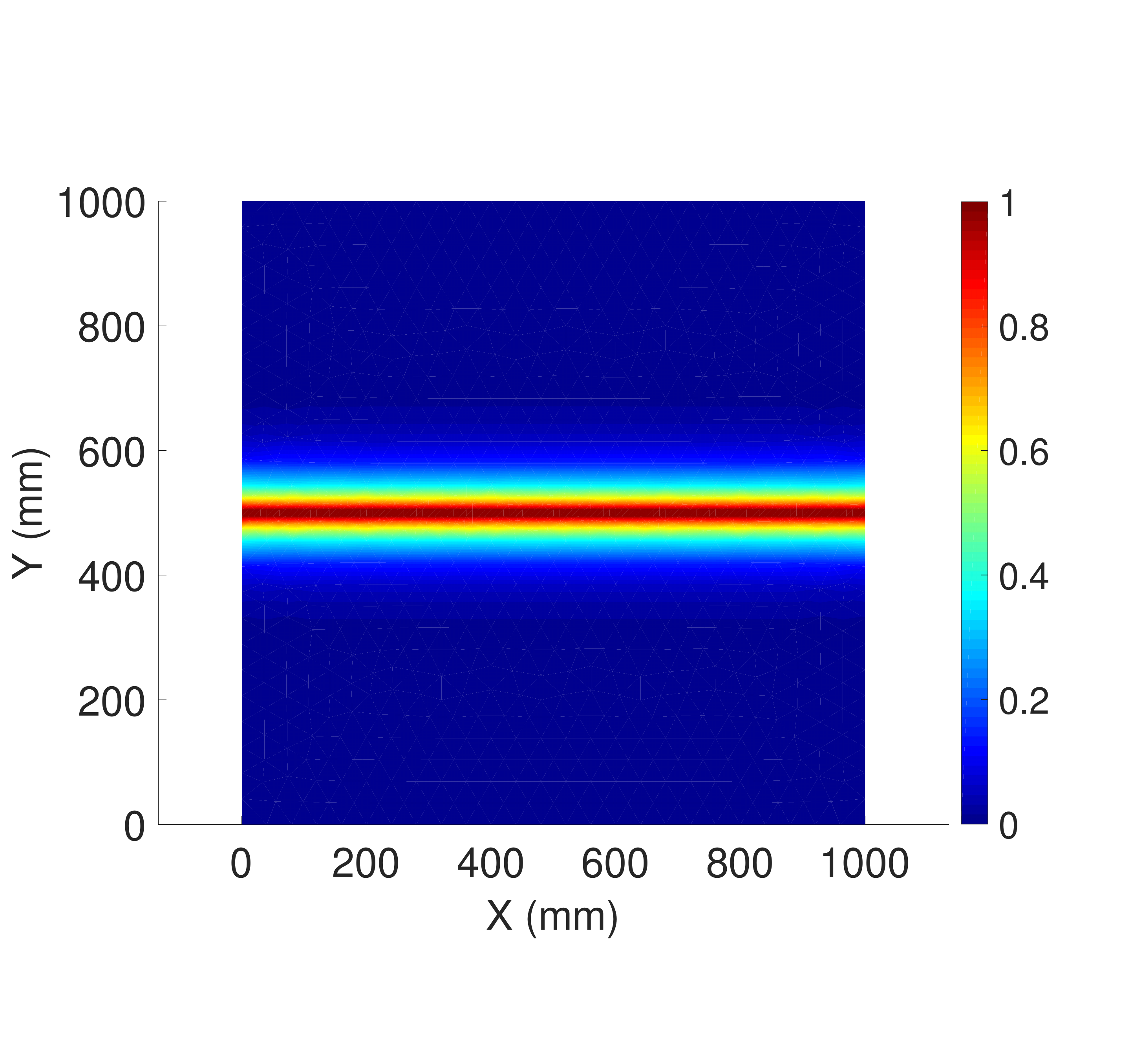}}
\quad
\subfloat[]{
\includegraphics[width=4in,height=2in,keepaspectratio]{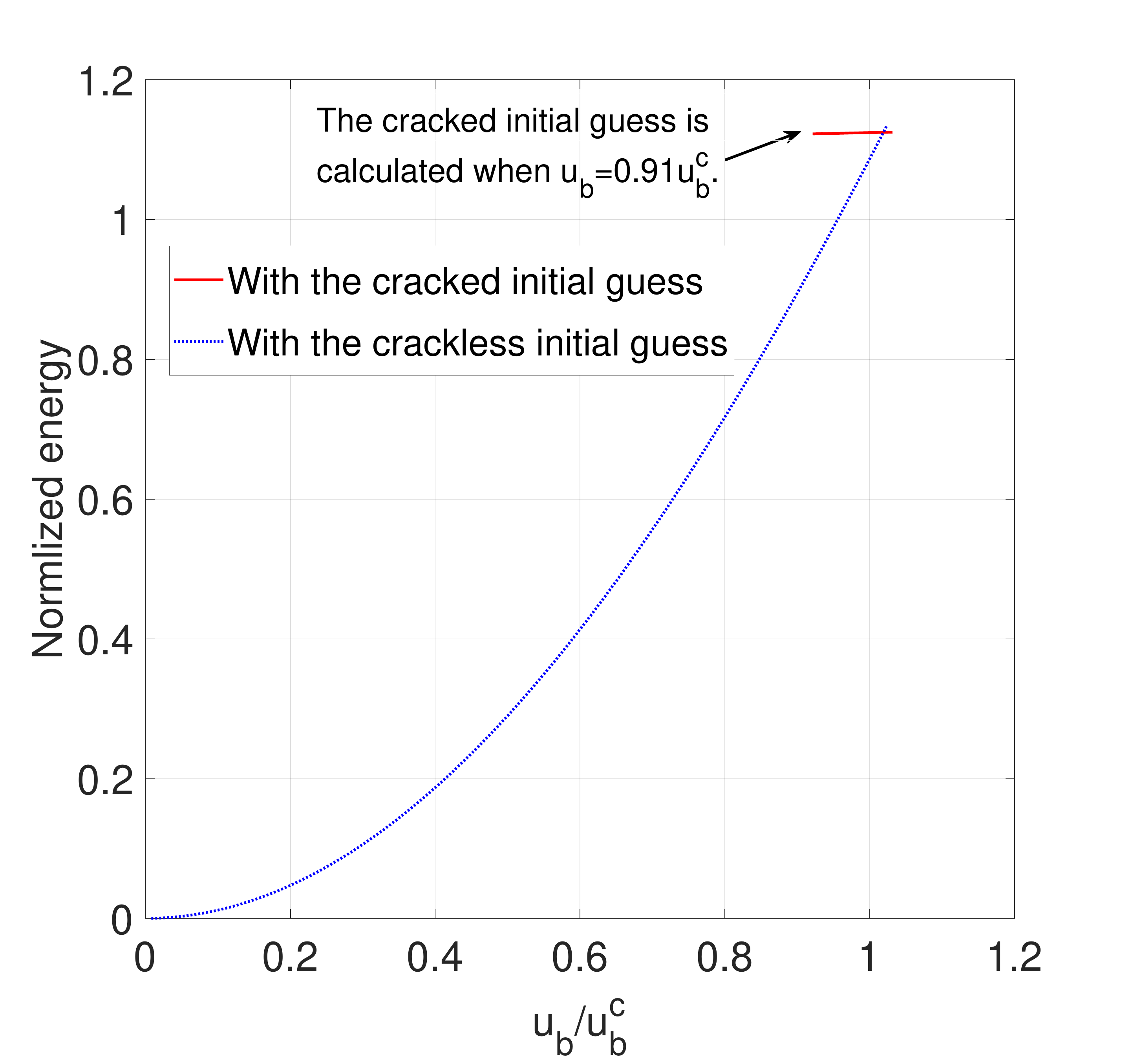}
\label{EnergyOfProposed}}
\quad
\subfloat[]{
\includegraphics[width=4in,height=2in,keepaspectratio]{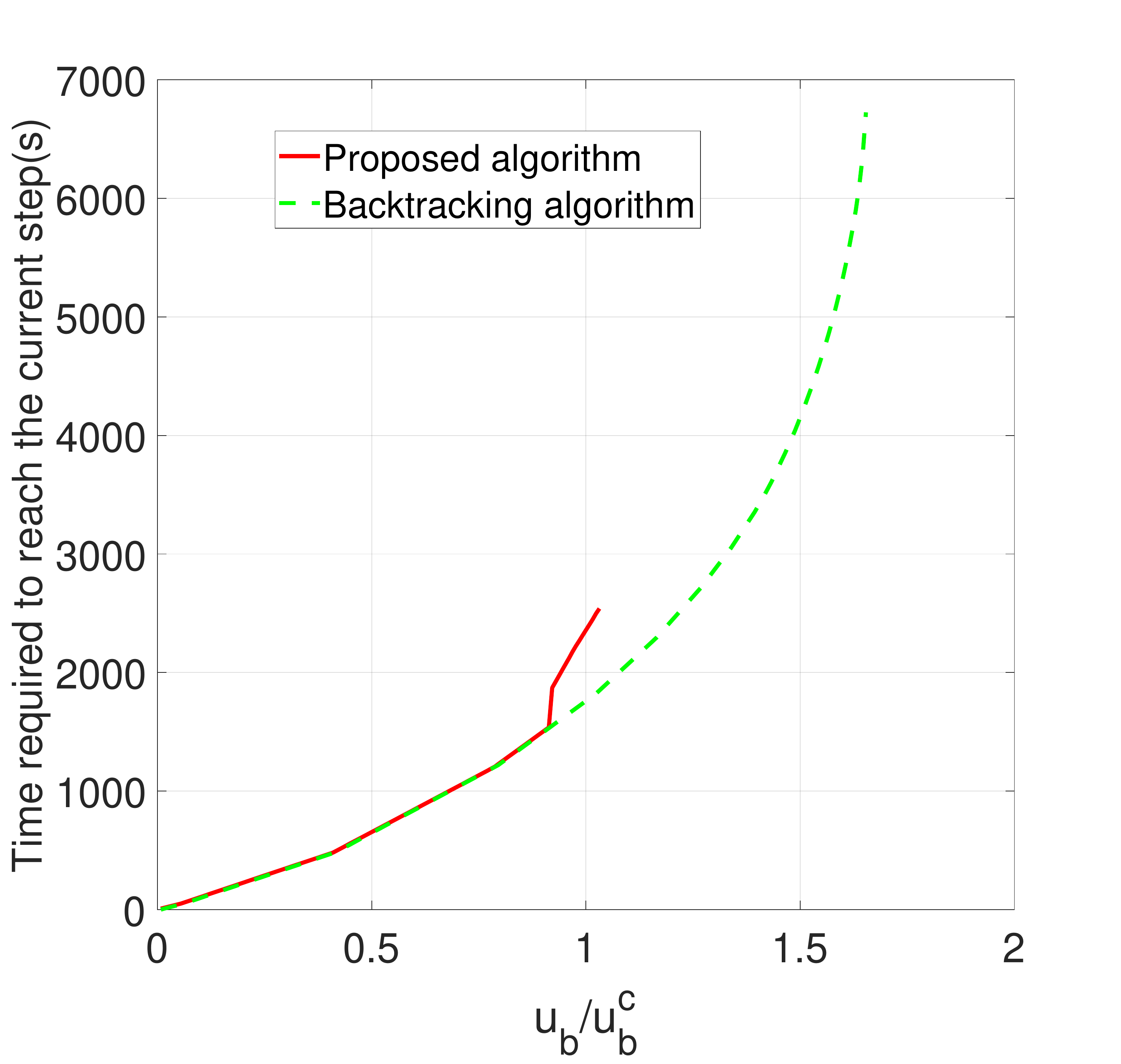}
\label{timeComparison}}

    \caption{Crack nucleation example of a homogeneous square. (a) Schematic. (b) Phase field of the cracked result at $u_b=u_b^c$. (c) Total energies of the proposed algorithm as a function of $u_b$,  with a crackless initial guess (blue) and the cracked initial guess (red) , the latter starts from when the load of vigilance $u_b^v=0.91u_b^c$ is reached. Here the energies are normalized by the total energy at complete fracture. (d) Comparison of the accumulated computational times between the proposed algorithm and the backtracking algorithm \cite{BOURDIN2007411} (only counting the first pass) with the same computational environment. When the backtracking algorithm passes the critical load $u_b^c$ for the first time, there is no warning until retraced from $u_b=1.66u_b^c$. Hence the proposed algorithm is much more cost effective from the beginning to fracture. }
    \label{energy Comparison}
\end{figure}

Next we consider a plane strain problem of a homogeneous square shown in Figure \ref{2dFigure2}. This example is challenging as there is no heterogeneity and no obvious site for crack nucleation. The initial domain is $\Omega=[-L/2,L/2]\times [-L/2,L/2]$. The boundary conditions on the upper and lower edges $(-L/2,L/2)\times\{\pm L/2\}$ are $(u_x,u_y)=(0,\pm u_b)$, where $u_{b}$ increases from zero quasistatically until the solid is completely fractured. The left and right edges $\{\pm L/2\}\times(-L/2,L/2)$ are traction free.  The parameters are listed in Table \ref{Tab:parameters3}.

\begin{table}[htbp]
	\centering
	\caption{Material parameters for Problem 3}
	\begin{tabular}{lccc}
		\toprule
		Parameter & Symbol & Value &Unit \\
		\hline
		Young's modulus & $E$ &210 &GPa \\
		Poisson's ratio & $\nu$ & 0.3 &-- \\
		Critical energy release rate & $G_c$ & 6750 & N/m \\
		Phase field length scale parameter & $\ell$ & 40 & mm \\
			Side length of the domain & $L$ & 1000 &mm\\
		\bottomrule
	\end{tabular}
\label{Tab:parameters3}
\end{table}

With symmetry, it can be shown that once a crack nucleates, it immediately propagates to completely fracture the solid. With this, a theoretical critical value for $u_b$, $u_b^c$, is given by

\begin{equation}
    u_b^c=\sqrt{\frac{G_{c}L}{2E}}.
\end{equation}

With numerical computation, the load of vigilance $u_b^v$ is found to be $0.91u_b^c$. After that parallel universes are initiated and the proposed algorithm starts to calculate the cracked initial guess and two converged candidate solutions with the cracked and crackless initial guesses. The cracked candidate is accepted for yielding a lower energy when the numerical critical value equal to $u_b=1.03u_b^c$ is reached, which is slightly larger than $u_b^c$. This discrepancy is known as the toughening effect due to finite element discretization \cite[Section 8.11]{Bourdin2008}.

We now compare the proposed algorithm with the standard Newton iteration and the backtracking method. First, With the standard Newton iteration, the crack is predicted to nucleate when $u_b=2.19u_b^c$, which is much higher than $u_b^c$.

The backtracking algorithm retraces $u_b=1.66u_b^c$ to $u_b=1.03u_b^c$ and then converges to a cracked result. In other words, the cracking load and results predicted by the two methods are the same. However, like in previous examples, the backtracking algorithm misses the possibility of fracture in the first pass, while the proposed algorithm does not, hence the latter is more reliable.

In terms of computation time, the backtracking algorithm costs 6725s while the proposed algorithm costs 2716s, which is 2.4 times faster.

\subsection*{Example 4: Anti-plane shear experiment on an anisotropic material}
\begin{figure}[htbp]
    \centering
    \subfloat[Schematic]{
\includegraphics[width=2in,keepaspectratio]{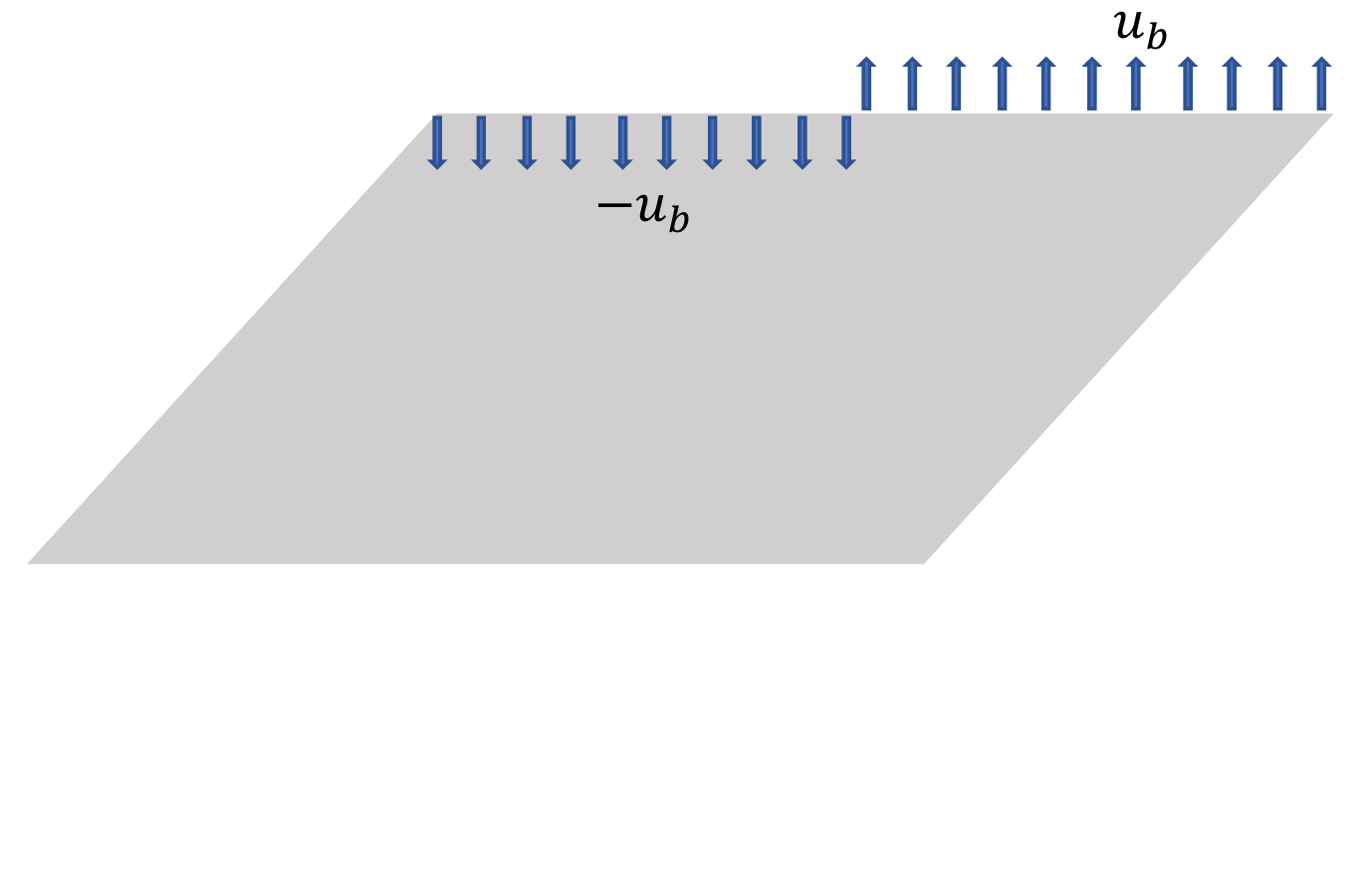}
\label{fig:Example4Config}
}
\quad
\subfloat[$\varepsilon=0.2$, $\beta=-45\degree$. $G_{c\rm{min}}=0.8$, $G_{c\rm{max}}=1.2$.]{
\includegraphics[width=2in,keepaspectratio]{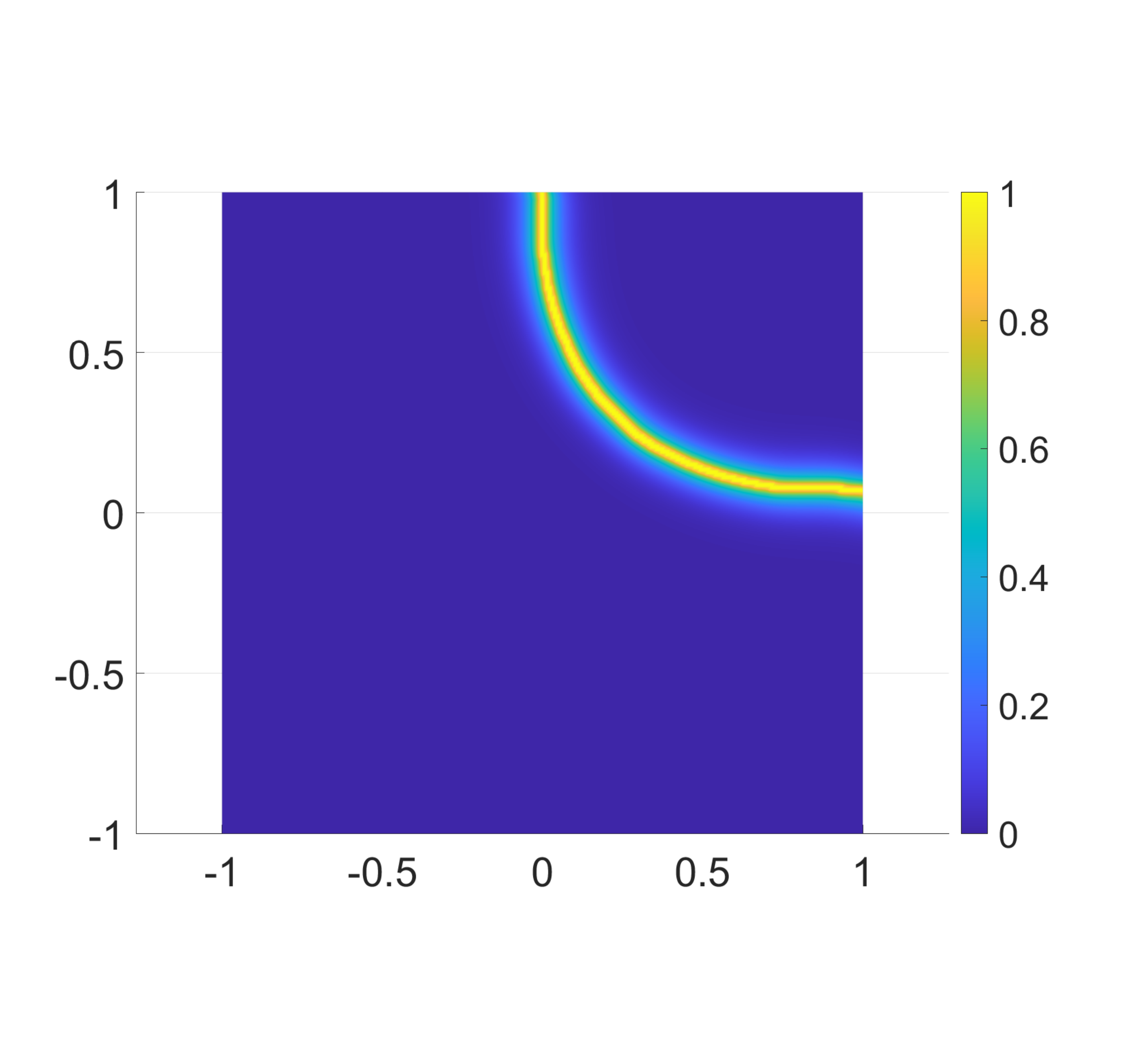}
}
\quad
\subfloat[$\varepsilon =0.2$, $\beta=-22.5\degree$. $G_{c\rm{min}}=0.8$, $G_{c\rm{max}}=1.2$. ]{
\includegraphics[width=2in,keepaspectratio]{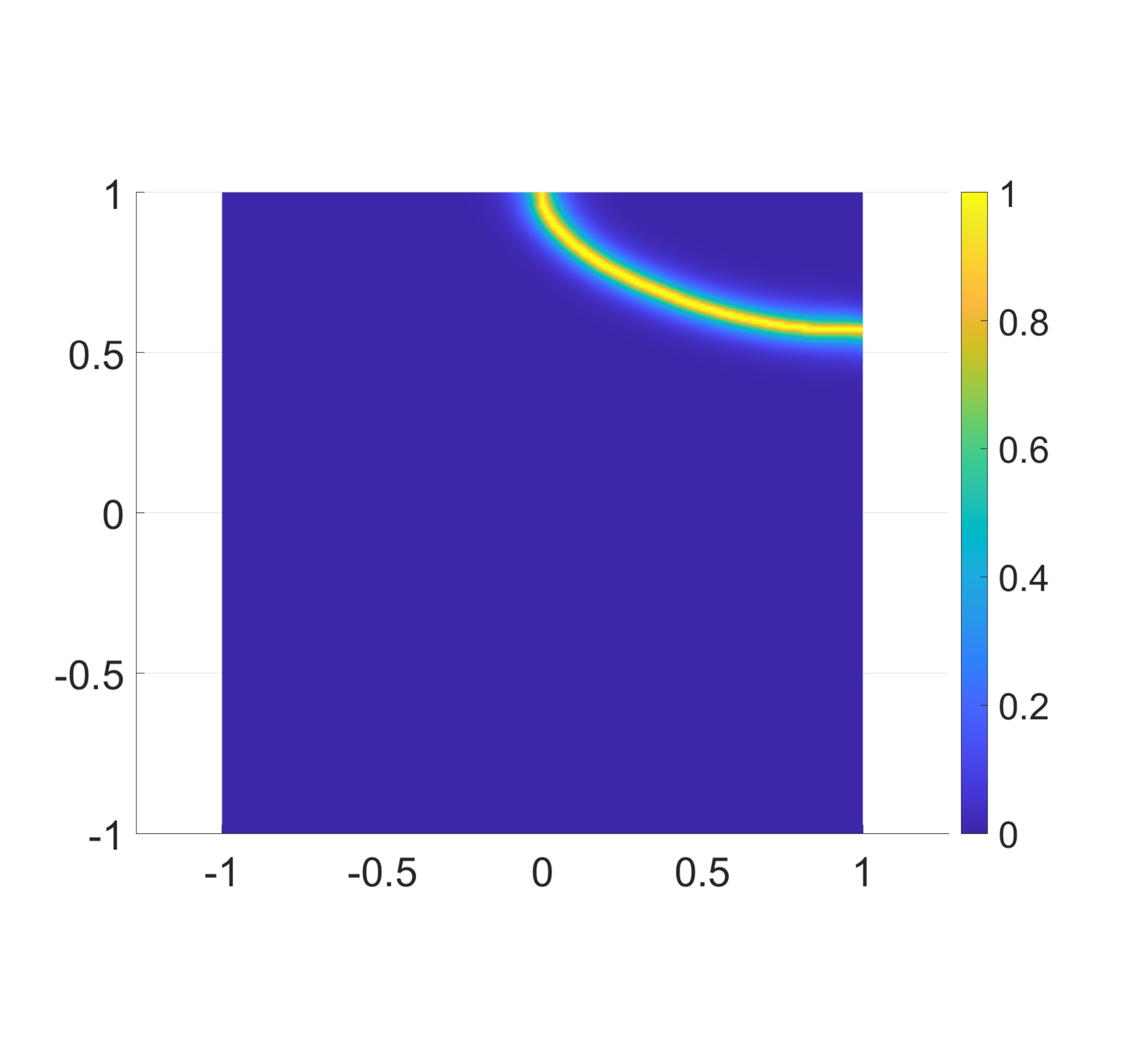}
}
\quad
\subfloat[$\varepsilon =0.2$, $\beta=-67.5\degree$. $G_{c\rm{min}}=0.8$, $G_{c\rm{max}}=1.2$. ]{
\includegraphics[width=2in,keepaspectratio]{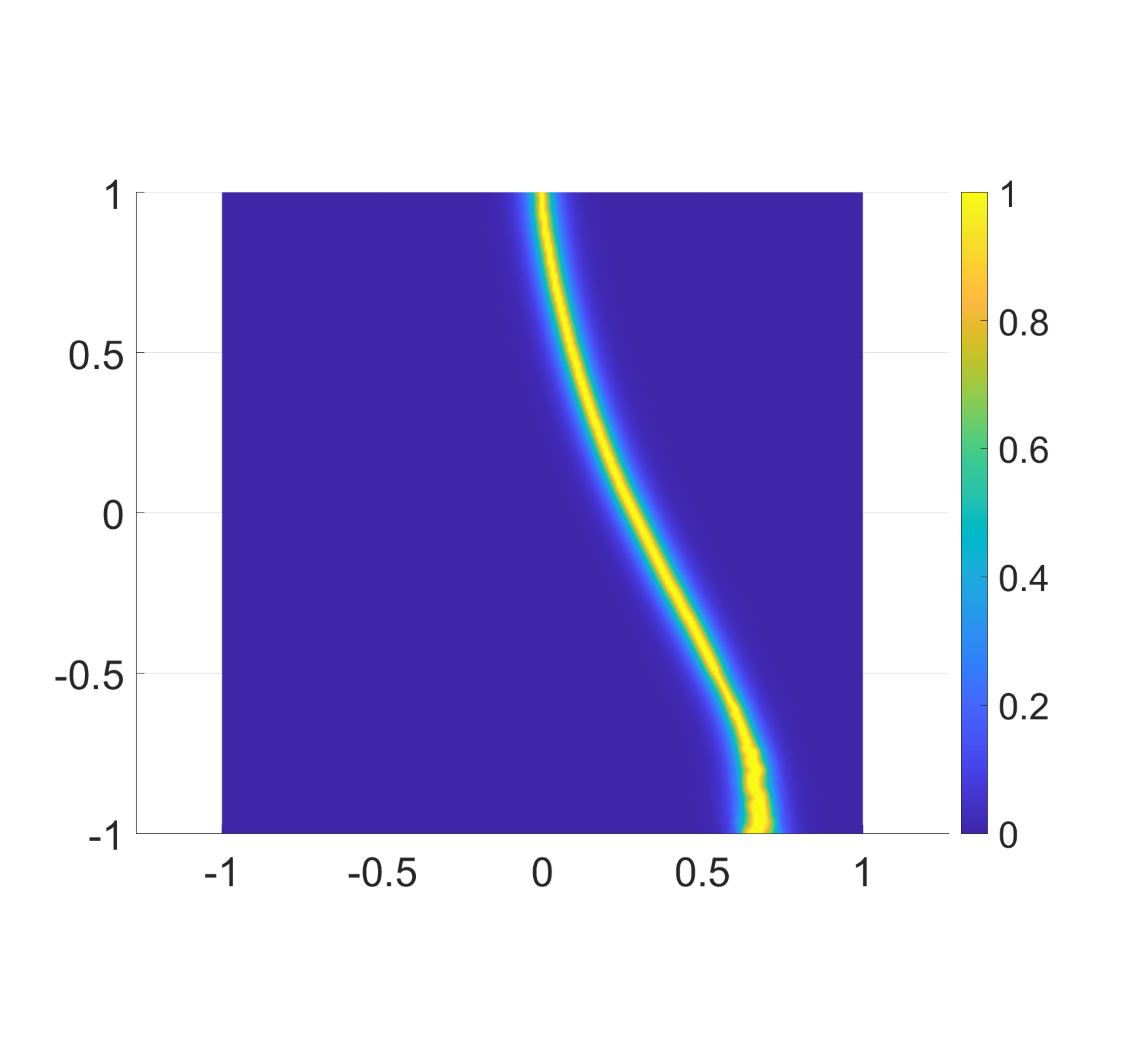}
}
\quad
\subfloat[$\varepsilon =0.5$, $\beta=-45\degree$. $G_{c\rm{min}}=0.5$, $G_{c\rm{max}}=1.5$.]{
\includegraphics[width=2in,keepaspectratio]{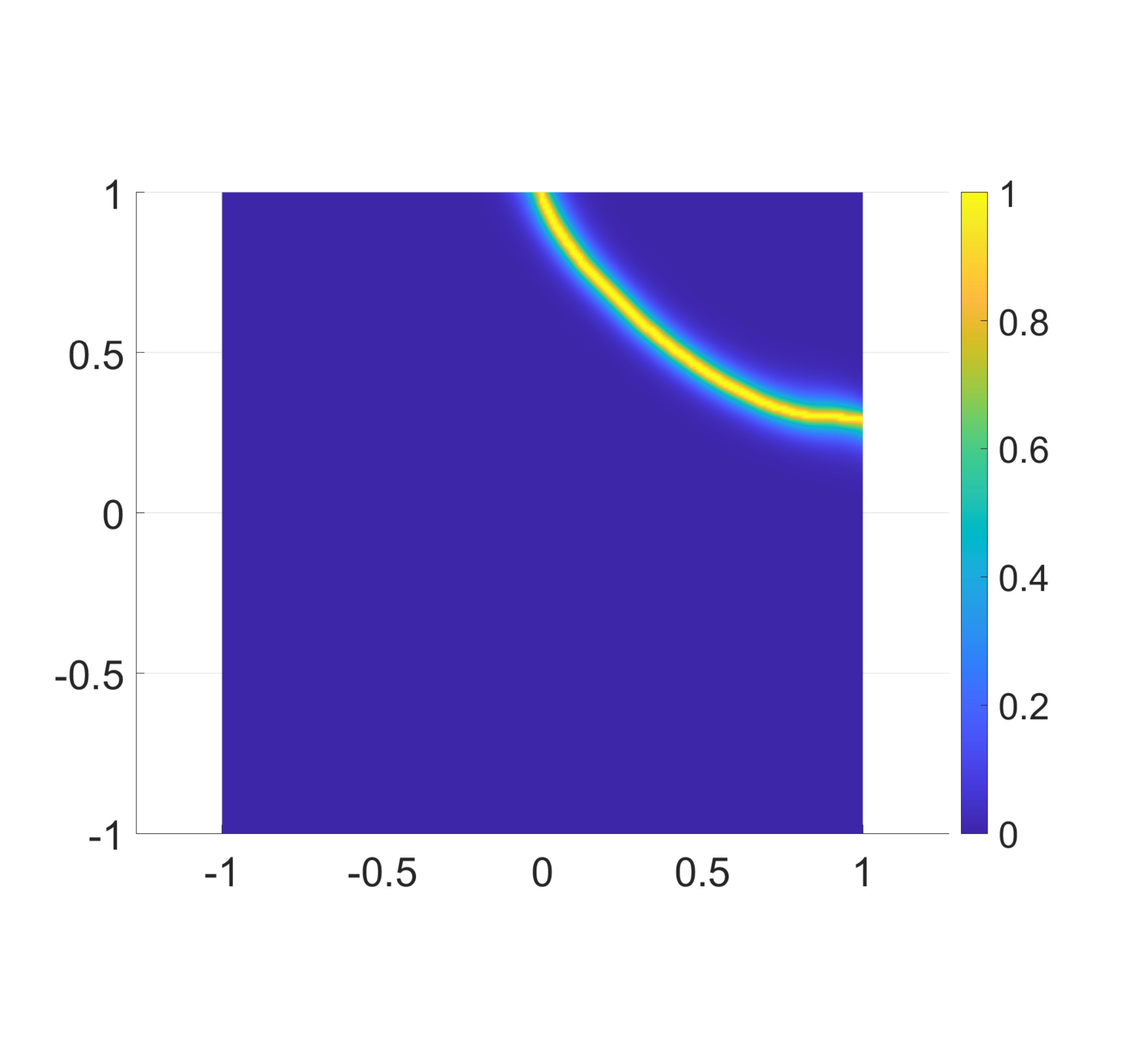}
}
\quad
\subfloat[$\varepsilon =0.8$, $\beta=-45\degree$. $G_{c\rm{min}}=0.2$, $G_{c\rm{max}}=1.8$.]{
\includegraphics[width=2in,keepaspectratio]{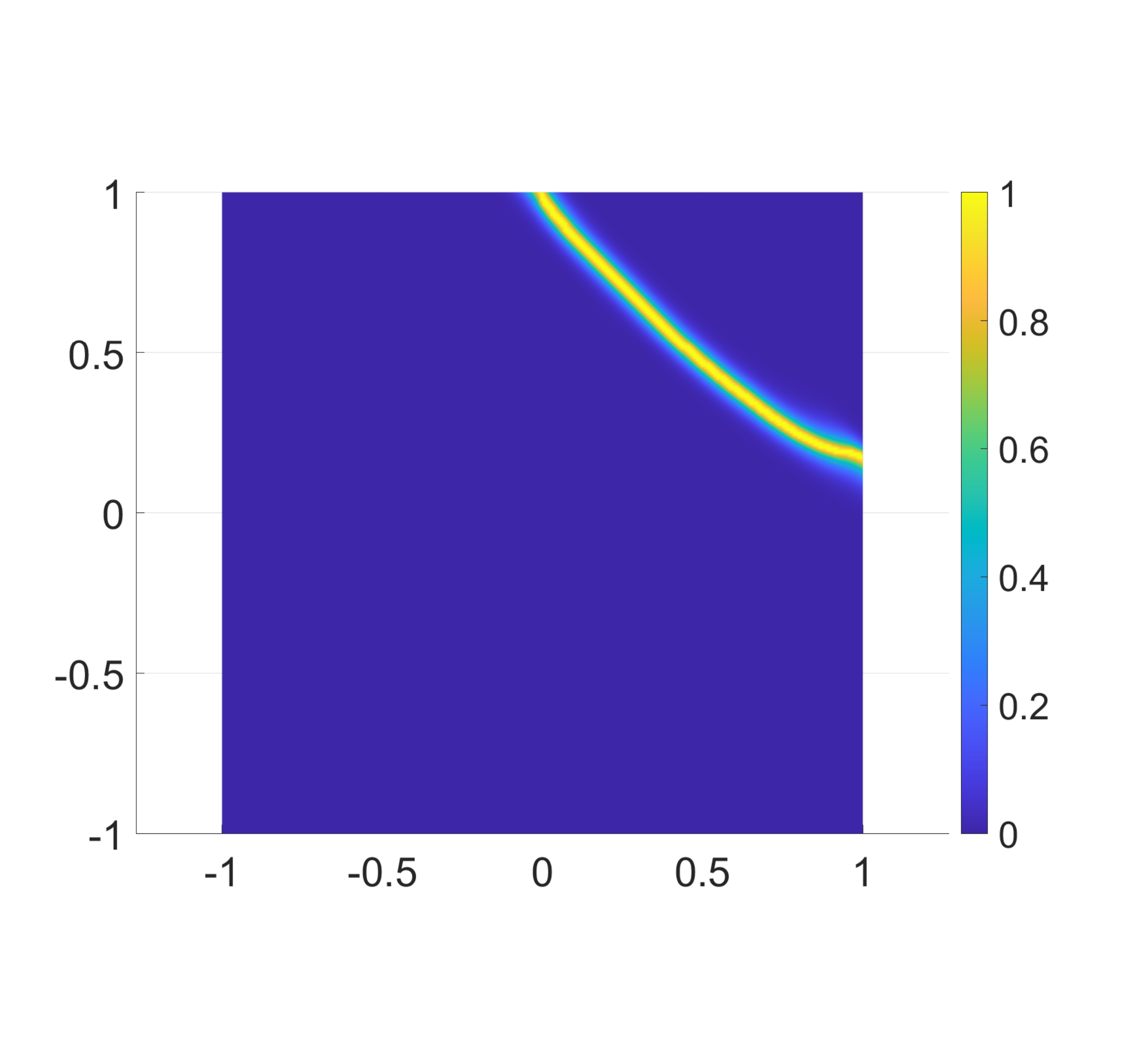}
}
    \caption{Anti-plane shear experiment on an anisotropic material.(a) Schematic. (b)-(f) plot the phase field of the result with parameters shown in each subtitle. As expected, the critical load of the proposed algorithm $u_b^{c}$ is smaller than that of the traditional Newton method $u_b^{c,N}$.  }
    \label{fig:Example4}
\end{figure}

We continue with an anti-plane tear experiment. As shown in Figure \ref{fig:Example4Config}, the domain is $(-L/2,L/2)\times(-L/2,L/2)$, where $L=2$. The boundary conditions on the upper edge $(-L/2,0)\times\{L/2\}$ and $(0,L/2)\times\{L/2\}$ are $u_z=\pm u_b$ respectively, where $u_{b}$ increases from zero quasistatically until the solid is completely fractured. The other edges are traction free. In this numerical example, the critical energy release rate $G_c$ is a function of fracture angle $\theta$, namely:
\begin{equation}
    G_c=G_{c}^{0}\left[1-\varepsilon \cos{(2(\theta-\beta))} \right]
    \label{GcAnisotropic}
\end{equation}
where $G_{c}^0$ is the average critical energy release rate, $\beta$ is the weakest material angle and $\varepsilon$ is the anisotropy strength, $0\leq\varepsilon\le1$. Then the minimum and maximum critical energy release rate $G_{c\rm{min}}=(1-\varepsilon)G_{c}^{0}$ and $G_{c\rm{max}}=(1+\varepsilon) G_{c}^{0}$, respectively. 

Equation \eqref{GcAnisotropic} is realized by replacing the second term of \eqref{Energy equation} by
\begin{equation*}
\int_{\Omega} G_c\left(\frac{d^{2}+\ell^{2}\left| \nabla d\right|^{2}+\ell^2\varepsilon\left[\cos{(2\beta)}(d_x^2-d_y^2)-2\sin{(2\beta)}d_x d_y\right]}{\mathrm{2} \ell}\right)\mathrm{d}\Omega,
\end{equation*}
where $d_x=\partial d/\partial x$ and $d_y=\partial d/\partial y$.

The parameters for the simulation are listed in Table \ref{Tab:parameters4}.

\begin{table}[htbp]
	\centering
	\caption{Material parameters for problem 4}
	\begin{tabular}{lccc}
		\toprule
		Parameter & Symbol & Value \\
		\hline
		Shear modulus& $\mu$ &1\\
		Average critical energy release rate & $ G_{c}^{0}$ & 1 \\
		Phase field length scale parameter & $\ell$ & 0.04 \\
		\bottomrule
	\end{tabular}
\label{Tab:parameters4}
\end{table}

For this problem, the load of vigilance $u_b^v$ is negligibly small.
 Figure \ref{fig:Example4} (b) through (f) plot the phase field of the results. Because of the competition of the strain energy and the fracture energy, the fracture direction lies roughly between the the normal direction of the edges and the direction with the minimum $G_{c}$. The crack starts and ends in approximately the normal direction of the edges. In the intermediate progress, the crack propagates in the direction with minimum $G_{c}$. This pattern is similar to the results reported in \cite{GERASIMOV2022114403}, in which a pre-existing crack was introduced.

Figure \ref{fig:Example4} (b), (c) and (d) show the crack paths with  $\beta$ increases.
Figure \ref{fig:Example4} (b), (e) and (f) show the crack paths 
  as $\varepsilon$ increases with $G_c^0$ fixed. Herein
  the crack path is more inclined in the direction with minimum $G_{c}$ as $\varepsilon$ increases. 
  
  As expected, the critical load from the proposed algorithm $u_b^{c}$, which is also the load at complete fracture, is smaller than that from the standard Newton iteration, $u_b^{c,N}$, see Table \ref{Tab:Example4Result}. This example also demonstrates the necessity of the proposed algorithm for crack nucleation.

\begin{table}[htbp]
	\centering
	\begin{tabular}{lccccc}
		\toprule
		$\varepsilon$ & $G_{c\rm{min}}$ & $G_{c\rm{max}}$ &$\beta$&$u_b^{c}$&$u_b^{c,N}$ \\
		\hline
		0.2 & 0.8 & 1.2 &$-45\degree$&  0.9&1.5\\
		0.2 & 0.8 & 1.2 &$-22.5\degree$&  1.0&1.1\\
		0.2 & 0.8 & 1.2 &$-67.5\degree$& 1.1&1.8 \\
	    0.5 & 0.5 & 1.5 &$-45\degree$&    0.9&1.1\\
	    0.8 & 0.2 & 1.8 &$-45\degree$&  0.3&1.2\\
		\bottomrule
		
	\end{tabular}
	\caption{Critical load of the proposed algorithm $u_b^{c}$ and that of the standard Newton iteration $u_b^{c,N}$.}
\label{Tab:Example4Result}
\end{table}

\section{Discussions}
\label{sec:Disccusion}
The proposed ``parallel universe'' algorithm provides an efficient means for crack nucleation problems in the context of the phase field approach to fracture. Both this algorithm and Bourdin's backtracking algorithm \cite{BOURDIN2007411} provide the same critical load for cracking and the same crack path, avoiding overestimating the strength of the material as opposed to the case of the standard Newton iteration. 

However, the proposed method is more reliable, as the backtracking algorithm normally requires computing the solution far beyond the desired load range in order to retrace the solution when cracking first appears.
In contrast, the proposed algorithm does not need such over computation, at the price of finding a (premature)  cracked candidate solution and of parallel computation of crackless and cracked candidate solutions until the cracked one is energetically favored. 

Moreover, the proposed algorithm requires shorter overall computational time, despite the doubling of the computation once the load of vigilance is reached.


A comment on the applicability of the method follows. As mentioned in the previous Section \ref{sec:Method}, up to now we have assumed $u_b^v\le u_b^c$, i.e., when the load of vigilance is reached, the energy with the cracked candidate solution is higher than the crackless counterpart, in which case the parallel universe computation just needs to be performed onward.
To make sure this equality holds, consider a square domain with length $L$ and a measure of stress concentration $K\ge1$ ($K=1$ for the homogeneous square domain). Displacement load $u$ and $-u$ are applied on the upper and lower boundary, respectively. Then the maximum principal stress is: 
\begin{equation*}
    \sigma_{\max}=\frac{2EKu}{L}.
\end{equation*}
Let $\sigma_{\max}=\sigma_{v}$, we get

\begin{equation*}
    u_b^v=\sqrt{\frac{27L^2G_{\rm{c}}}{1024\alpha^2 EK^2\ell(1-\nu^{2})}}.
\end{equation*}

A simple calculation of $u_b^c$:
\begin{equation*}
    u_b^c=\sqrt{\frac{G_{\rm{c}}L}{2E}}.
\end{equation*}

Then a sufficient condition for this inequality is given by the ratio of $L/\ell$:
\begin{equation}
\frac{L}{\ell}\le 
\frac{512(1-\nu^2)K^2\alpha^2}{27}, \label{applicability}    
\end{equation}

Equation \eqref{applicability} may be the basis for generalizing the proposed scheme to the case of domain decomposition, with $L$ the characteristic domain size.


Although the current version of the proposed algorithm applies only to brittle quasi-static fracture, generalization to the cases of elastoplastic fracture and dynamic fracture is straightforward.

\section*{Acknowledgments}
We acknowledge the financial support by the National Natural Science Foundation of China, Grant No.~11972227, and by the Natural Science Foundation of
Shanghai, Grant No.~19ZR1424200.


\end{document}